\documentclass[trackchanges,twocolumn]{aastex62}

\usepackage{blindtext}
\usepackage{textcomp}
\usepackage{afterpage}

\newcommand{\sao}{\affiliation{Smithsonian Astrophysical Observatory, Cambridge, MA, USA}}
\newcommand{\umich}{\affiliation{University of Michigan, Ann Arbor, MI, USA}}

\newcommand{\ucb}{\affiliation{University of California, Berkeley, CA, USA}}

\newcommand{\baja}{\affiliation{Baja Technology LLC, Tempe, AZ, USA}}
\newcommand{\iowa}{\affiliation{Department of Physics and Astronomy, University of Iowa, Iowa City, Iowa, USA}}
\newcommand{\lpp}{\affiliation{Laboratoire de Physique des Plasmas, CNRS, Sorbonne Universite, Ecole Polytechnique, Observatoire de Paris, Universite Paris-Saclay, Paris, 75005 France}}
\newcommand{\gsfc}{\affiliation{Heliophysics Science Division, NASA, Goddard Space Flight Center, Greenbelt, MD 20771, USA}}
\newcommand{\uofa}{\affiliation{Lunar and Planetary Laboratory, University of Arizona, Tucson, AZ 85721, USA}}

\graphicspath{{./}{figures/}}

\received{}
\revised{}
\accepted{}

\submitjournal{ApJ (in work)}

\shorttitle{The Parker Solar Probe SPAN-Electron Sensors}
\shortauthors{Whittlesey et al.}

\begin{document}
\title{The Solar Probe ANalyzers - Electrons on Parker Solar Probe}

\correspondingauthor{Phyllis L. Whittlesey}
\email{phyllisw@berkeley.edu}

\author[0000-0002-7287-5098]{Phyllis L. Whittlesey}\ucb
\author[0000-0001-5030-6030]{Davin E. Larson}\ucb
\author[0000-0002-7077-930X]{Justin C. Kasper}\umich\sao
\author[0000-0001-5258-6128]{Jasper Halekas}\iowa
\author{Mamuda Abatcha}\ucb
\author{Robert Abiad}\ucb
\author[0000-0001-6235-5382]{M. Berthomier}\lpp
\author[0000-0002-3520-4041]{A. W. Case}\sao
\author{Jianxin Chen}\baja
\author{David W. Curtis}\ucb
\author{Gregory Dalton}\ucb
\author[0000-0001-6038-1923]{Kristopher G. Klein}\uofa
\author[0000-0001-6095-2490]{Kelly E. Korreck}\sao
\author[0000-0002-0396-0547]{Roberto Livi}\ucb
\author{Michael Ludlam}\ucb
\author{Mario Marckwordt}\ucb
\author{Ali Rahmati}\ucb
\author{Miles Robinson}\ucb
\author{Amanda Slagle}\ucb
\author[0000-0002-7728-0085]{M. L. Stevens}\sao
\author{Chris Tiu}\gsfc
\author[0000-0003-1138-652X]{J. L. Verniero}\ucb

\begin{abstract}
Electrostatic analyzers of different designs have been used since the earliest days of the space age, beginning with the very earliest solar wind measurements made by Mariner 2 en route to Venus in 1962. The Parker Solar Probe (PSP) mission, NASA's first dedicated mission to study the innermost reaches of the heliosphere, makes its thermal plasma measurements using a suite of instruments called the Solar Wind Electrons, Alphas, and Protons (SWEAP) investigation.  SWEAP's electron Parker Solar Probe Analyzer (SPAN-E) instruments are a pair of top-hat electrostatic analyzers on PSP that are capable of measuring the electron distribution function in the solar wind from 2 eV to 30 keV. For the first time, in-situ measurements of thermal electrons provided by SPAN-E will help reveal the heating and acceleration mechanisms driving the evolution of the solar wind at the points of acceleration and heating, closer than ever before to the Sun. This paper details the design of the SPAN-E sensors and their operation, data formats, and measurement caveats from Parker Solar Probe's first two close encounters with the Sun.
\end{abstract}
\keywords{plasmas, space vehicles: instruments, solar wind, Sun: corona}

\section{Introduction} \label{sec:intro}

\begin{figure*}[ht!]
\centering
\includegraphics[height=6.5cm]{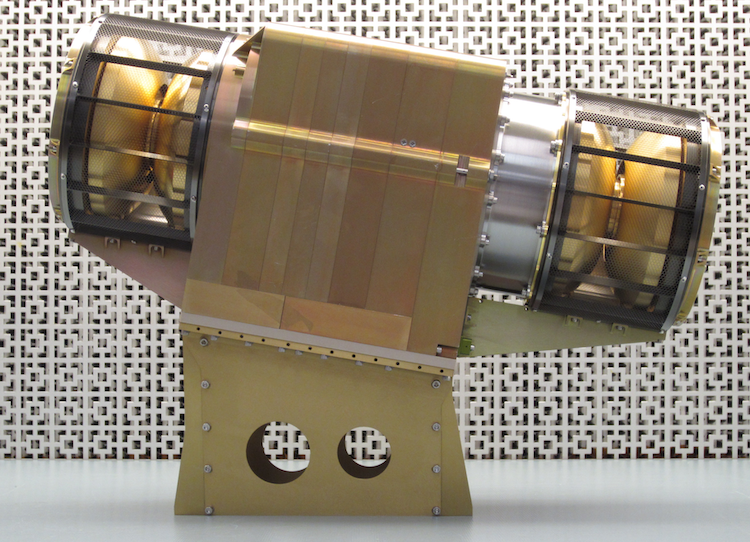} \includegraphics[height=6.5cm]{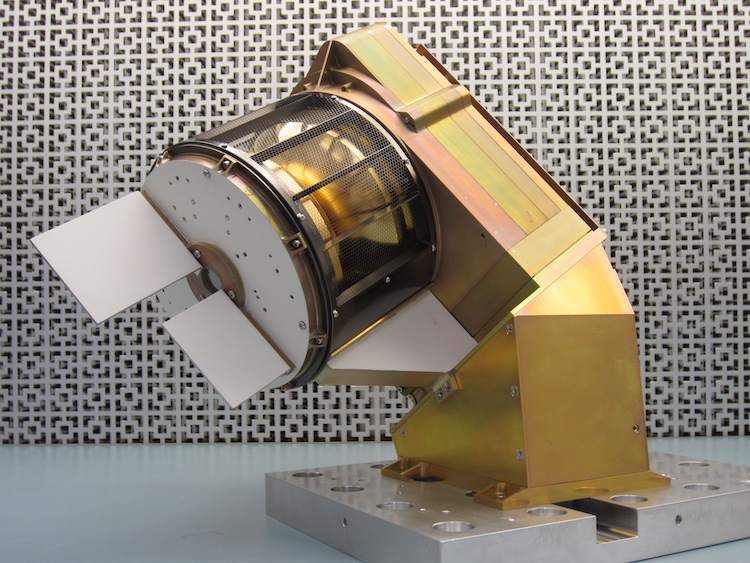}
\caption{Left, the SPAN-A composite instrument, consisting of two ESA instruments, SPAN-Ion and SPAN-Ae. The sensor on the left is the SPAN-Ae electron sensor, and the sensor to the right is the SPAN-Ion instrument. The deflectors, the polished gold-colored surfaces in the analyzer heads, are clearly visible in the image. Right: The SPAN-B instrument, mounted to a thick aluminum plate (non-flight). SPAN-B was designed to be safe to operate when in sunlight, and part of that design includes the use of radiator fins on top of the analyzer (shown) and a radiator plate under the electronics box (obscured in figure) on in combination with AZ-2000-IECW, which functions as a radiative coating.   \label{fig:photoSPANS}}
\end{figure*}
 
\subsection{Spacecraft and Suite}
Parker Solar Probe (PSP) is a NASA robotic mission flying closer to the Sun than any previous spacecraft. 
PSP is a three-axis-stabilized, primarily Sun-pointed spacecraft in a highly elliptical heliocentric orbit, with aphelia between the Earth and Venus. 
Using multiple Venus gravity assists between 2018 August and 2025 March, PSP's perihelion distance is incrementally lowered from 35 solar radii (R$_S$) at the start of the mission to 9.86 solar radii at the end of the prime mission. 
PSP's overarching science questions are: 1) Determine the structure and dynamics of the magnetic fields at the sources of the fast and slow solar wind; 2) Trace the flow of energy that heats the solar corona and accelerates the solar wind; and 3) Explore mechanisms that accelerate and transport energetic particles. PSP uses an encounter-based operations scheme: the science instruments on the spacecraft collect their primary, high cadence data during a solar ``encounter'' phase that lasts for 10-15 days (when the distance from the Sun is less than $0.25$ AU) around perihelion, and occasionally collect lower cadence data during the remaining portion of the orbit, termed the ``cruise'' phase. More details on the mission architecture and science goals can be found in \citet{Fox2015}.

 The ``Solar Wind Electrons, Alphas, and Protons'' (SWEAP) investigation is the thermal energy particle-detecting instrument package on PSP. To achieve the mission's science goals, SWEAP characterizes the bulk plasma in the solar wind and corona by measuring  the low-energy (\textless30 keV) ion and electron populations. SWEAP consists of 3 Electrostatic Analyzer instruments, called the Solar Probe ANalyzers (SPANs), and the Solar Probe Cup (SPC), all controlled by the SWEAP Electronics Module (SWEM), a SWEAP-specific instrument digital processing unit. On the ram side of the spacecraft one electron SPAN sensor, called SPAN-Ae, and one ion SPAN sensor, called SPAN-I or SPAN-Ion, are mounted as part of a single package, and collectively are called SPAN-A. On the anti-ram side of the spacecraft a single SPAN instrument is mounted to measure electrons, called SPAN-B. The top-level objectives and overall design of the SWEAP suite have been described previously in \citet{Kasper2016}. The individual SPAN-A and SPAN-B sensor packages are shown in Figure \ref{fig:photoSPANS}.

\startlongtable
\begin{deluxetable*}{c c c c c}
\tablecaption{The components within the SWEAP suite \label{tab:sweap_instruments}}
\tablehead{
\colhead{Name} & \colhead{Type} & \colhead{Particle Measured} & \colhead{Measurement Type} & \colhead{Look Direction}
}
\startdata
SPAN-I & Electrostatic Analyzer + ToF	& Ions		                & 3D VDF + mass             & Ram \\
SPAN-Ae & Electrostatic Analyzer 	    & Electrons		            & 3D VDF                    & Ram \\
SPAN-B  & Electrostatic Analyzer 	    & Electrons		            & 3D VDF                    & Anti-Ram \\
SPC     & Faraday Cup                   & Ions and Electrons		& 1D VDF + flow-direction   & Nadir\\
SWEM    & Instrument Digital Processing Unit & N/A & N/A & N/A
\enddata
\end{deluxetable*}
The two SPAN instruments that measure the 3-dimensional ($\theta$ and $\phi$ directions and energy) electron velocity distribution functions (VDF) by using concentric toroidal hemispheres for energy discrimination are collectively called SPAN-Electron, or SPAN-E for short. The SPAN-E sensors use multiple discrete anodes to measure incoming particles from different azimuth ($\phi$) angles. Electrostatic deflectors scan through multiple elevation ($\theta$) angles to measure incoming electron flow velocity directions. More details regarding Electrostatic Analyzer (ESA) design, construction, history, and operation can be found in \citet{Carlson1998} and \citet{Carlson1982}. ESAs of different designs have been measuring plasmas in space for six decades of exploration in the Earth's magnetosphere, ionosphere, the solar wind, the atmospheres of other planets, and at the boundaries of our solar system \citep[e.g.,][]{Mobius1998, Lin1995b, McFadden2008a, Neugebauer1997, Verscharen2019a}.

\begin{figure*}[ht!]
\includegraphics[height=8cm]{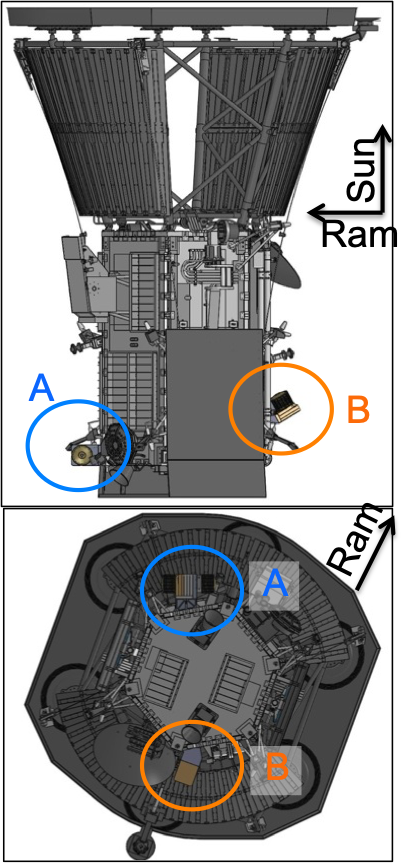}\includegraphics[height=8cm]{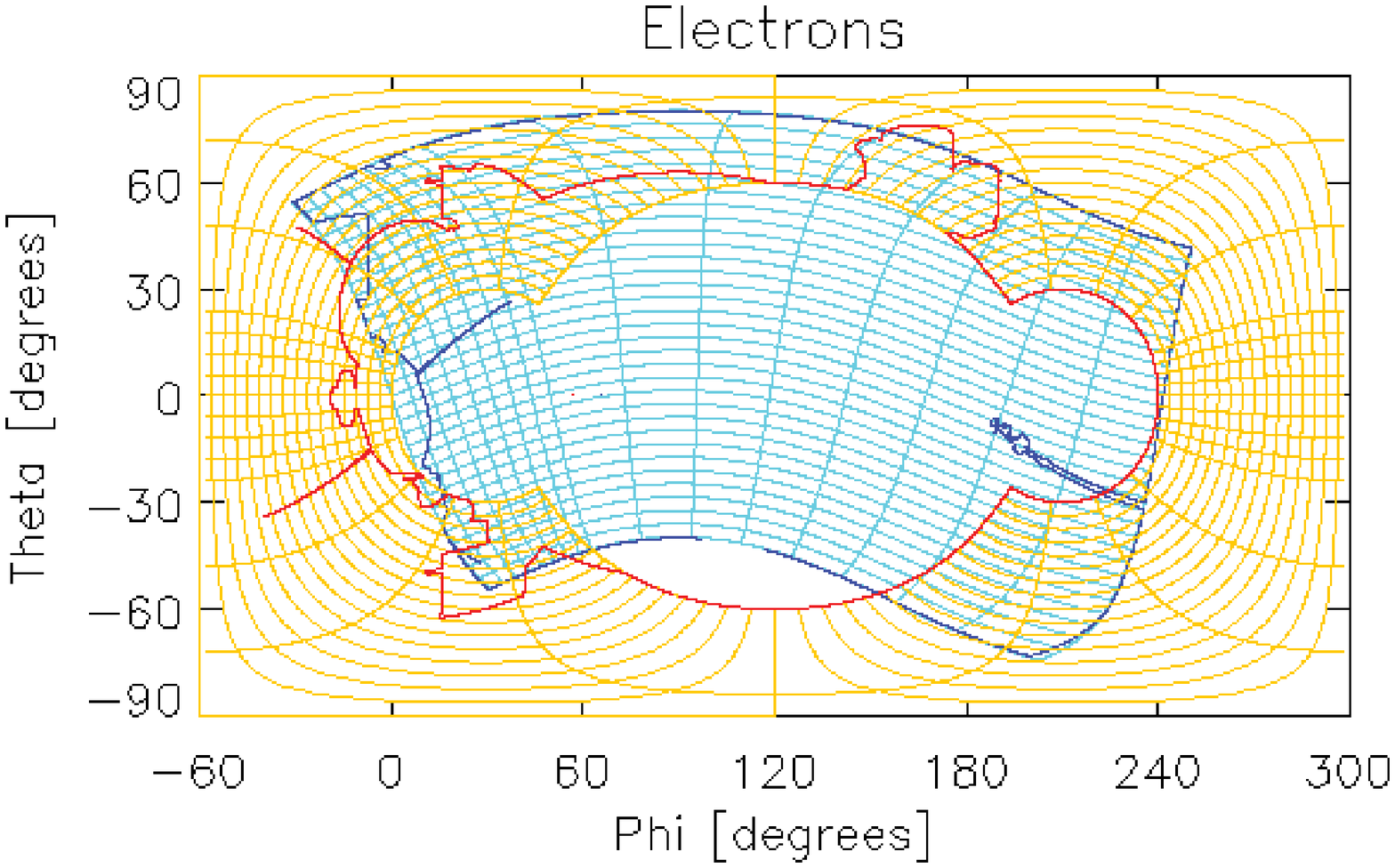}
\caption{The two SPAN-E Sensors' individual fields of view, overlaid on the same sky map. The blue trace is the extent of the SPAN-Ae FOV, and the orange is for the SPAN-B sensor. The red trace describes intrusions to SPAN-B's FOV by the spacecraft, and the dark blue trace, intrusions to SPAN-Ae. Identifiable features include the magnetic field instrument boom, two FIELDs antennas, and the high gain antenna. The Sun is at (0,0), and the ram-direction is at (90,0)
\label{fig:FOV}}
\end{figure*}

 The SPAN-Ion instrument differs from the SPAN-Electron sensors in that it contains a time-of-flight (ToF) section that permits identification of a particle's mass per unit charge, and also in that it measures positive ions and not electrons or negative ions. Further details on the SPAN-Ion instrument can be found in \citet{Livi}. 
 
 The SPC instrument is a Faraday Cup, which is a planar sensor consisting of a series of grounded grids along with a grid that is biased with an oscillating voltage that selects specific passbands of particle energy/charge, permitting their collection on four conducting plate segments at the back of the sensor. Details on SPC can be found in \citet{Case} as well as in \citet{Kasper2016}. 
 
 \begin{figure*}[ht!]
\centering
\includegraphics[height=6.5cm]{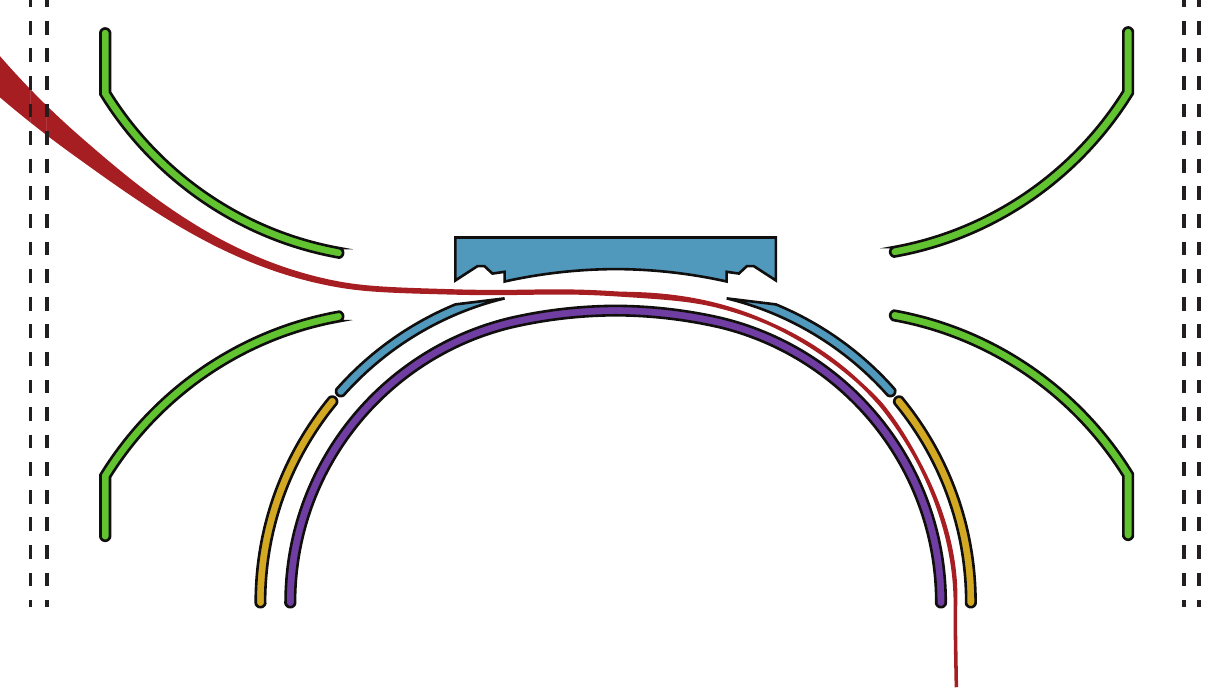} 
\caption{ Graphical cross-section of the SPAN-E optics design. Green lines represent the curves of the deflectors, blue represents the top cap and upper outer hemisphere, yellow is the spoiler section, and purple is the inner hemisphere. The red curve is a representative travel path for electrons eligible to pass through the analyzer when the hemisphere (purple) is biased, the upper deflector is biased, the lower deflector is at ground, and both the spoiler and upper outer hemisphere (blue and orange) are also held at ground.  \label{fig:optics_cross}}
\end{figure*}

 Lastly, the Solar Wind Electronics Module (SWEM) is the instrument computer that powers and controls all the SWEAP instruments. The SWEM relays commands from the ground/spacecraft to the instruments, stores command sequences onboard for single or repeated execution either by absolute time sequence (ATS) or via real-time commanding from the ground. The SWEM also monitors the current and voltage safety limits for the instruments, and stores high-resolution SPAN data onboard. PSP telemetry restrictions permitting, these data products are eventually downlinked for ground-based analysis. The entire suite of plasma instruments are summarized in Table \ref{tab:sweap_instruments}.

\subsection{The Electron Solar Probe ANalyzers}
The Electron Solar Probe ANalyzers (SPAN-E) are a pair of electrostatic analyzers that measure solar wind electrons from all sky directions that are unobstructed by other spacecraft components (e.g., the heat shield, solar panels, and the FIELDS antennas). The SPAN-Ae sensor is mounted on the ram side of the spacecraft, sharing a mounting point with SPAN-Ion. SPAN-B is mounted on the anti-ram side. SPAN-Ae and SPAN-B's individual fields-of-view (FOVs) fit together as shown in Figure \ref{fig:FOV}. A previous iteration of the SPAN-E instruments was detailed in \citet{Kasper2016}, but the final flown design and operational scheme from Encounters 1 \& 2 are detailed here. This paper exists to document the instruments' mechanical and electrical design, operational scheme, and data product format and contents for members of the scientific community intending to use the SPAN-E data. Since PSP is an encounter-based mission with progressively closer perihelia over time, the final calibration values and updates to those values will be detailed over the mission lifetime in future paper releases.

\section{Instrument Description} \label{sec:instrument_description}

The SPAN-E instrument design borrows heavily from previous ion and electron ESAs developed at the University of California, Berkeley for missions such as MAVEN/SWIA \citep{Halekas2015}, THEMIS/ESA  \citep{McFadden1998}, and Wind/3DP \citep{Lin1995b}. The SPAN-E optical design is based on heritage designs from MAVEN, which themselves improved on previous instruments cited above. The SPAN-E electronics depart substantially from previous designs, utilizing an ASIC (Application Specific Integrated Circuit) preamplifier chip and embedded capacitors instead of the discrete preamplifiers and capacitors used in heritage designs, a change which saves mass and avoids saturation effects.

\subsection{Basic Electrostatic Analyzer Operation}

Electrostatic analyzers select particles based on their energy/charge (E/q) through the use of nested, concentric curved surfaces; for the typical top-hat design, the surfaces are semi-hemispherical or semi-toroidal. For shorthand, both surfaces are often interchangeably referred to as "hemispheres", and this paper will follow that convention unless explicitly stated. The outer hemisphere contains an opening at the apex to serve as an aperture to allow particles to enter the space between the two hemispheres, which are indicated in Figure \ref{fig:optics_cross} by the purple, blue, and yellow surfaces. In the case of SPAN-E, the outer hemisphere is held at ground and the inner hemisphere is biased positively, creating an electric field between the two surfaces that permits electrons of a selected energy to pass through the curved surfaces without impacting the interior surfaces. By controlling the voltage bias on the inner hemisphere, the ESA selects a specific, narrow energy passband of particles, and particles outside of that passband impact the sides and are not measured. Narrower gaps between the two concentric hemispheres result in a smaller passband, or a smaller $\Delta E/E$, and an analyzer with higher energy resolution and lower geometric factor \citep{Carlson1998}.

The range of elevation angles over which incident particles are able to enter into the ESA opening aperture is narrow, typically about 5-15$^\circ$. The acceptance angle is intrinsically linked to the size of the gap between hemispheres. To increase the field of view, curved deflectors are added outside of the ESA hemisphere entrance aperture, as seen in the green surfaces in Figure \ref{fig:optics_cross}. These deflectors are biased positively to ``pull'' the electrons in from wider angles of incidence into the intrinsic field of view of the hemispheres. When one deflector is biased positively, the other is held at ground. Thus, by scanning through a range of deflector voltages, the analyzer field of view can be increased to up to 120$^\circ$. To prevent the deflectors from leaking electric field and disturbing the ambient plasma, a pair of grounded grids are fixed around the analyzer outside of the deflectors (see Figure \ref{fig:block_diagram} for position of ground grids relative to deflector). 

 When an electron successfully passes through the deflectors and the ESA hemispheres, it exits the hemispheres at an angle of incidence normal to the microchannel plates and anode board underneath the hemispheres. A grid at the exit aperture of the hemispheres is held at ground to prevent stray electric fields from leaking out. Micro-channel plates (MCPs) are used to convert a single incident charged particle into an electron shower of millions of particles. The MCPs themselves are biased such that electrons are accelerated through the plates and onto an anode board, where they are counted as a single electron ``count'' on metallized pads that correspond to the phi direction pixel size in the data. 
 
\startlongtable
\begin{deluxetable*}{l l l}
\tablecaption{SPAN-E Instrument Design Parameters
\label{tab:instrument_params}}
\tablehead{\colhead{Parameter} & \colhead{Value\tablenotemark{a}} & \colhead{Comments}}
\startdata
Analyzer Radii                & R1 = 3.34 cm                              & Inner Hemisphere Toroidal Radius\\
                              & R2 = R1 * 1.03 = 3.440 cm                 & Outer Hemisphere Toroidal Radius\\
                              & R3 = R1 * 1.639 = 5.474 cm                & Inner Hemisphere Spherical Radius\\
                              & R4 = R3 * 1.06 = 5.803 cm                 & Outer Hemisphere Spherical Radius \\
                              & RD = 3.863 cm                             & Deflector Spherical Radius \\
R2 - R1 separation            & 1 mm                                      & Toroidal Top-Hat  \\
Opening Angle Hemisphere      & 13\textdegree\                            & \\
Opening Angle Top Cap         & 12\textdegree\                            & \\
Analyzer Constant             & 16.7                                      & As derived from the optics model \\
Analyzer voltage (max)        & 0 V to 2000 V                             & Controllable to less than a Volt \\
Deflector Voltage             & 0 V to 4000 V                             & Controllable to less than a Volt \\
Spoiler Voltage               & 0 V to 80 V                               & set to zero by default (no attenuation)\\
Energy Range                  & 2eV to 30keV                              & \\
Analyzer energy resolution    & 7\%                                       & \\
Spoiler Attenuation Factor    & 8\tablenotemark{b}                        & setting for E1 \& E2; varies w/ energy channel \\
Energy sweep rate             & 32 steps in 0.218 sec\tablenotemark{c}    & \\
Deflector sweep rate          & 8 steps / 32 ``microsteps" in 6.80 ms     & ``microsteps'' in full sweeps only \\
Spoiler sweep rate            & 32 steps in 0.218 sec                     & Only when enabled - zero by default \\
Azimuth range                 & 240\textdegree                            & \\
Instantaneous field of view   & 240\textdegree $\times$ 3.507\textdegree  & $\theta$ = 0\textdegree (no deflection) \\
Field of view each sweep      & 240\textdegree $\times$ 120\textdegree    & FOV blockage varies by sensor\\
Anode angle resolution        & 6\textdegree\ or 24\textdegree            & 8 ``small" anodes and 8 ``large" \\ 
Analyzer geometric factor     & 0.00101 cm$^2$ sr E                       & prediction from simulations, 240\textdegree analyzer only\\
                              & 5.984 $\times$ 10$^{-4}$ cm$^2$ sr E      & with 5 $\times$ 90\% transparency grids\tablenotemark{d}\\
Measurement Cadence           & 0.435 sec                                 & For either Full or Targeted Sweeps (not both)\\
Measurement Duration          & 0.218 sec                                 & for 32 energy by 8 deflector bins \\
Counter readout               & 0.852 ms                                  & 32 energy by 8 deflector bins per sweep
\enddata
\tablenotetext{a}{Note that values in the above table are as designed values; final calibrated values to be included in a future SPAN-E calibration paper.}
\tablenotetext{b}{Estimated, final calibration pending spoiler use in Encounter 3 and beyond.}
\tablenotetext{c}{Note that any combination of energy and deflection bins so long as their product is equal to 256.}
\tablenotetext{d}{Not including MCP efficiency}

\end{deluxetable*}
 
 By varying the voltages on the deflectors and hemisphere as a function of time, the time-ordered train of positive electron detections at time $t$ can be sorted by deflector and hemisphere voltage, and thus a distribution of electrons as a function of energy (hemisphere voltage) and direction (deflector voltage and anode number) can be created.


\subsection{Optical Design: Toroidal Hemispheres}\label{sec:optical_design}

The optical design in SPAN-E is a toroidal top-hat analyzer design, with deflectors permitting a FOV in the $\theta$ direction of $\pm$ 60\textdegree. The anode board defines the $\phi$ direction, and in the case of SPAN-E anodes are laid out over 240\textdegree. The design parameters used to create the electron sensor optics are detailed in Table \ref{tab:instrument_params}. A narrow gap between the hemispheres was chosen in order to achieve a higher energy resolution (a lower $\Delta E/E$) and a lower sensitivity to prevent saturation close to the Sun. Picking a narrow gap while maintaining the same size analyzer means that the geometric factor of the instrument is reduced, which helps offset the chance for saturating since the electron densities are higher in the inner heliosphere compared to those at 1AU.


\subsection{Optical Design: Deflectors}\label{sec:deflectors}

Both SPAN-Ae \& SPAN-B are in the shadow of the heat shield during normal encounter spacecraft science operations, so generation of photoelectrons off of illuminated surfaces of the SPAN-E sensors is not a usual concern. All exposed SPAN-E surfaces are operated at positive voltage biases. The deflectors themselves (see Figure \ref{fig:photoSPANS}) are also mounted inside cylindrical, grounded grids across the instrument aperture of 92\% open transparency each to prevent stray electric field from the deflectors escaping and contaminating the ambient plasma. The aperture ground grids also affect the angular response of the instrument in the extreme $\theta$ look directions.

During periods when the spacecraft is slewed so the anti-ram side is sunlit (e.g., outside 0.7 AU and during communications with Earth), SPAN-B is illuminated. Since the exposed biased surfaces of SPAN-B are only ever biased positive or set to zero, and since under nominal conditions a sunlit spacecraft has a positive potential, they are not typically a driver of spacecraft potential.


\subsection{Attenuation Methods}\label{sec:attenuators}

Another way in which the SPAN-E design deviates from heritage design is the methods of attenuation. Much like MAVEN/SWIA, MAVEN/STATIC, and THEMIS/SST, SPAN-E has a mechanical attenuator, consisting of a thin metal piece of perforated metal which can be moved into the instrument aperture when the particle number flux exceeds a certain threshold. The mechanical attenuator is designed to be used when the particle count rate nears the intrinsic sensor saturation \citep{Wilson2011}. The mechanical attenuator in SPAN-E consists of a series of slits aligned with the instrument's axis of symmetry that reduce the total particle flux by a factor of 10. 

However, to accommodate the wide range of fluxes expected over the duration of the mission, another factor of attenuation was added to the SPAN-E instrument design. The outer hemisphere of SPAN-E is split into two components, an upper hemisphere and a lower hemisphere, seen as the blue and yellow curves in Figure \ref{fig:optics_cross}. The upper-outer hemisphere (blue) is held at ground, as is typical of ESA design. The lower-outer hemisphere (yellow), called the ``spoiler'', is either commanded to ground, or swept between a voltage of 0-80V max at the same cadence as the inner hemisphere. When the spoiler is at ground, the upper and lower parts of the outer hemisphere are at the same value and the analyzer operates as a typical ESA. When the spoiler is swept, analyzer optics are ``spoiled", effectively narrowing the energy passband (reducing the $\Delta E/E$) and thereby reducing the geometric factor of the instrument. Thus, the total number of particles permitted through the hemispheres is reduced, and the electron flux through the instrument drops. The factor of attenuation is a function of the ratio of the commanded spoiler voltage to the commanded inner hemisphere voltage, and in the lab was capable of reducing the count rate to effectively zero particles per second, indistinguishable from the MCP background noise. During typical on-orbit operations when electrostatic attenuation is enabled, the SPAN-E spoiler primarily attenuates electrons with energies less than 500eV. Other methods of electrostatic attenuation that have been used on past ESA instruments are detailed in \citet{Collinson2010}. 

 
\subsection{Instrument Mechanical Design}\label{sec:mechanicalDesign}

The two SPAN-E sensors are positioned on the spacecraft in such a way that the two fields of view combine together to cover $>90\%$ of the sky, as shown in Figure \ref{fig:FOV}. The SPAN-Ae sensor is mounted on the ram side of the spacecraft near the bottom deck. The SPAN-Ae FOV in the $\phi$ coordinate is defined as beginning 6\textdegree\ from the sunward direction when the spacecraft is sun-pointed, and extends roughly through the ecliptic plane around the back of the spacecraft by 240\textdegree. The deflectors extend the intrinsic 3.5\textdegree\ FOV above and below the ecliptic plane by 60\textdegree\ on either side, defining the $\theta$ angle. Eight 6\textdegree -wide anodes span the 6\textdegree - 54\textdegree $\phi$ angles, or the angles defined by the plane of the anode board. The remaining $\phi$ angles are filled with anodes 24\textdegree\ wide. See Figure \ref{fig:FOV} for the mounting point of SPAN-Ae and how the intrinsic angular resolution maps onto the sky. 

\begin{figure*}[ht!]
\centering
\includegraphics[height = 6cm]{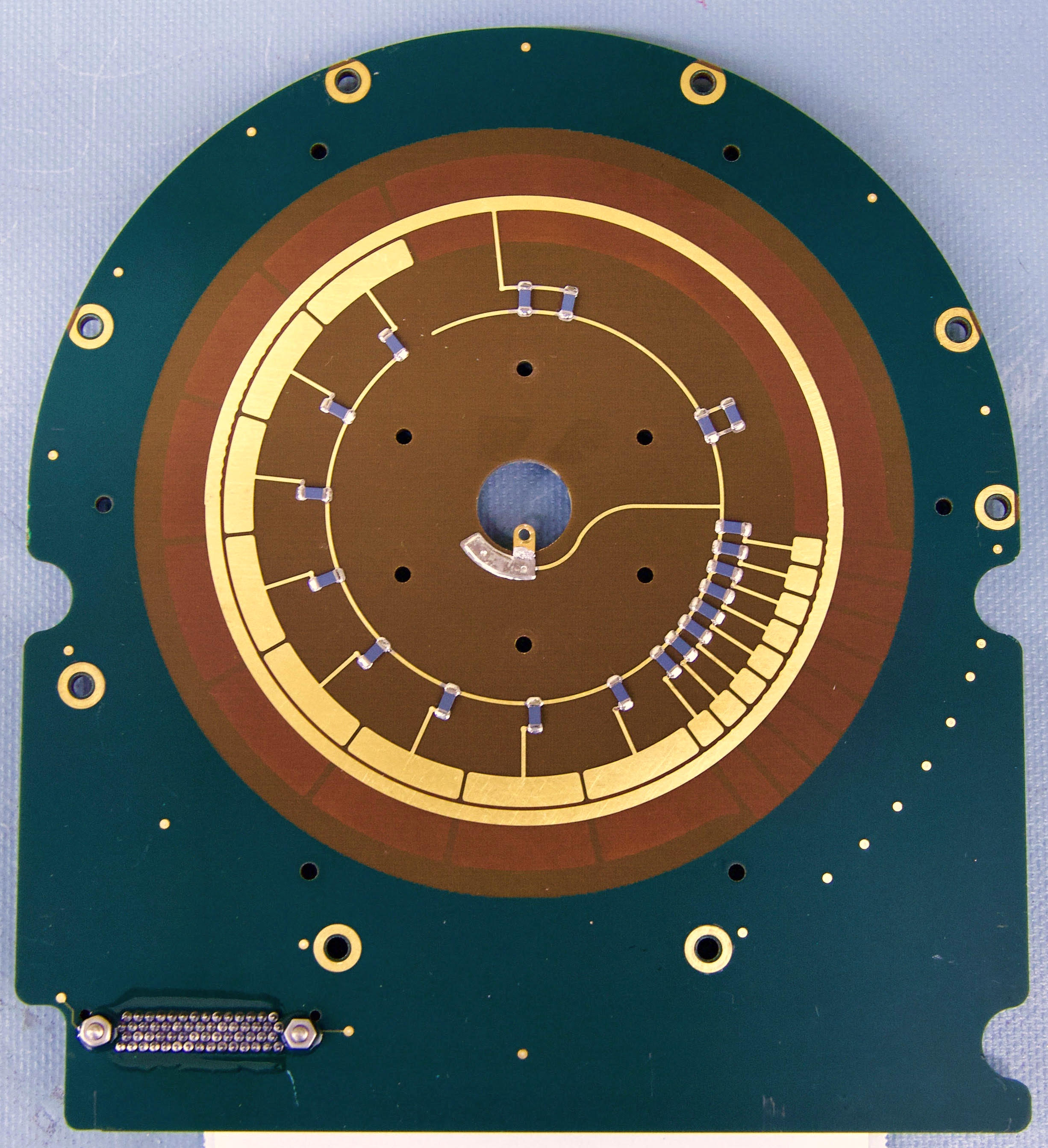}\includegraphics[height = 6cm]{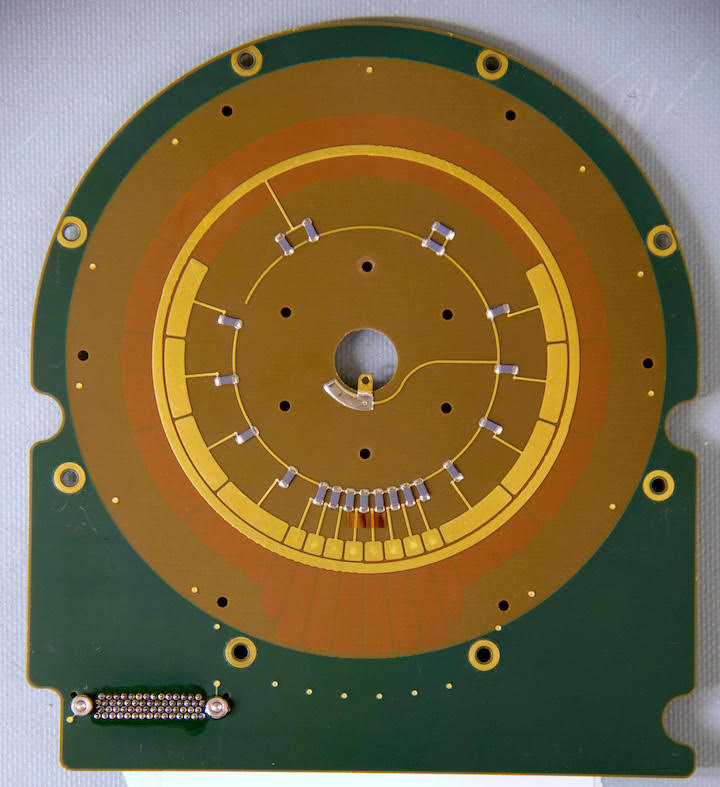}
\caption{\textbf{Left:} the SPAN-Ae Anode board, with anodes visible in the 3:00 to 11:00 positions. Small anodes are seen from roughly 3:00 to 5:00, and correspond to SPAN-Ae look directions in the ecliptic plane near the Sun direction. \textbf{Right:} the SPAN-B anode board, with anodes visible from roughly 2:00 to 10:00. Small anodes are seen from roughly 5:00 to 7:00, and correspond to a FOV segment perpendicular to and bisecting the ecliptic plane on the anti-ram side of the spacecraft.
\label{fig:bothAnodeBoards}}
\end{figure*}

SPAN-B, on the other hand, is mounted on the anti-ram side of the spacecraft. Where SPAN-Ae's $\phi$ coordinate is parallel to roughly the ecliptic plane, the SPAN-B $\phi$ coordinate, or the plane of the anode board, is roughly orthogonal to the ecliptic plane, and slightly tilted toward the Sun. The SPAN-B anode board clusters its small 6\textdegree anodes closer to where the $\phi$ plane bisects the ecliptic plane (120\textdegree\ above and below the ecliptic plane), and since the 240\textdegree\ FOV of SPAN-B is centered on the ecliptic plane, they span $\pm$ 24\textdegree above and below the ecliptic. The remaining $\phi$ angles are filled with 24\textdegree wide anodes. The SPAN-B deflectors also increase the FOV to +\- 60\textdegree\ about the $\phi = 0$ plane. Figure \ref{fig:bothAnodeBoards} contains an image of the SPAN-Ae and SPAN-B anode boards, which indicate the pixel sizes and orientations for both instruments.

\section{SPAN-E Electronics}\label{sec:measurement_electronics}

\begin{figure*}[ht!]
\centering
\includegraphics[scale = 0.6]{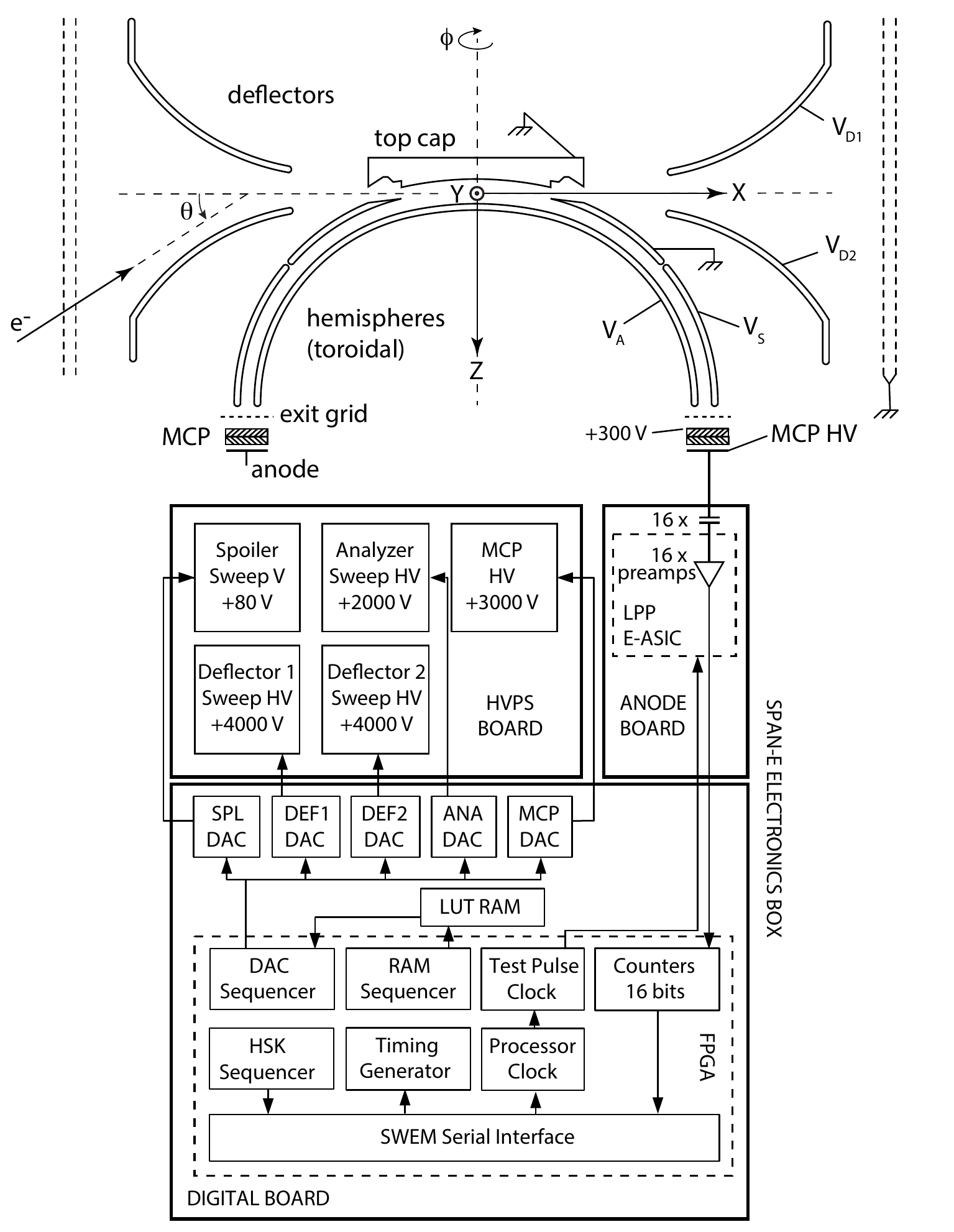}
\caption{A block diagram of a SPAN-E instrument that has been simplified to its core measurement components. The bolded dashed lines in the center of the analyzer cross-section reflect the symmetry in hemispheres and deflectors that correspond to the phi and theta instrument coordinates. $V_A$ represents the Analyzer bias, which determines the electron energy; $V_{D1}$ \& $V_{D2}$ represent the two deflector biases. $V_S$ represents the spoiler voltage.
\label{fig:block_diagram}}
\end{figure*}

\begin{figure*}[ht!]
\centering

\includegraphics[height=5cm]{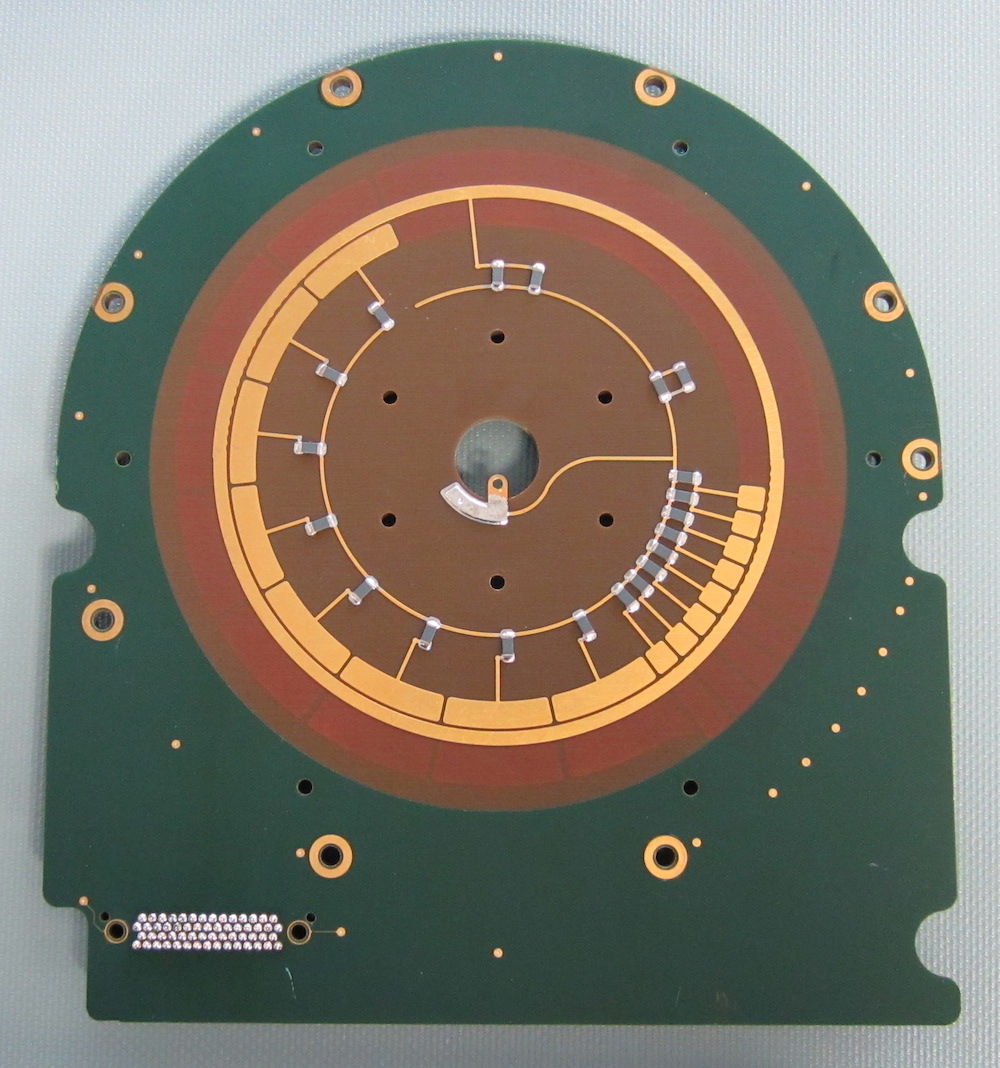} \includegraphics[height=5cm]{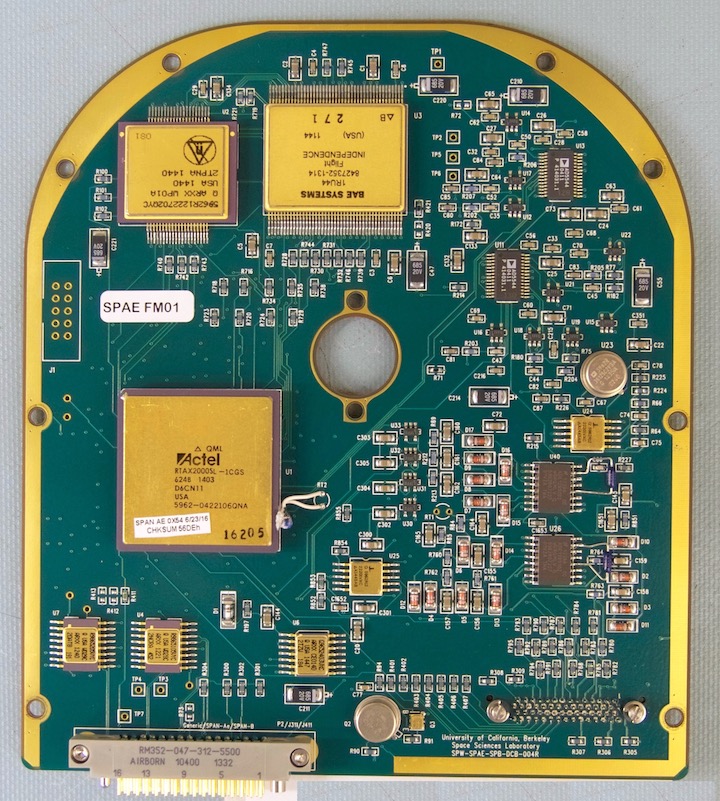} \includegraphics[height=5cm]{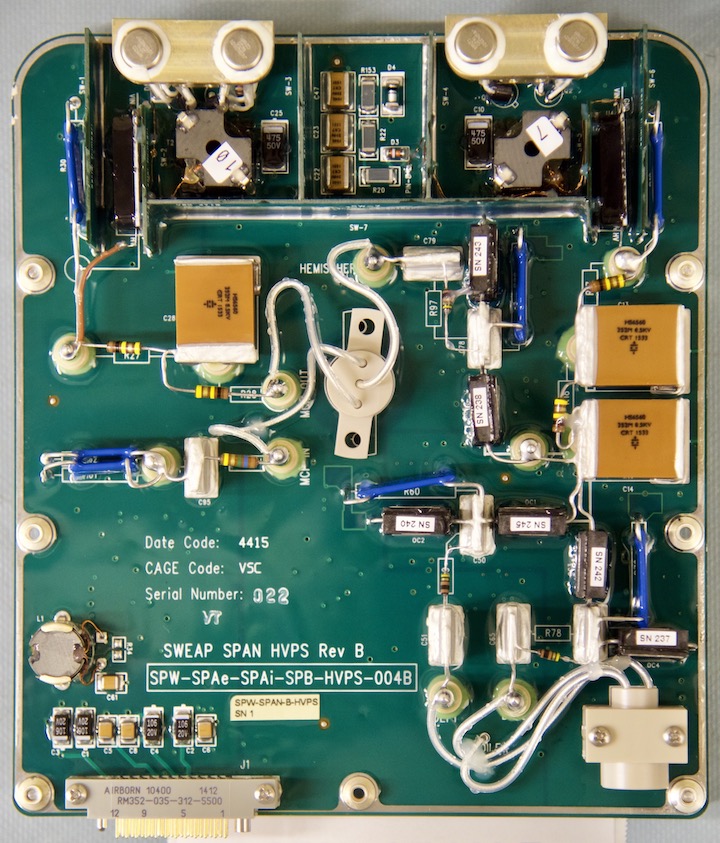} 
\includegraphics[height=5cm]{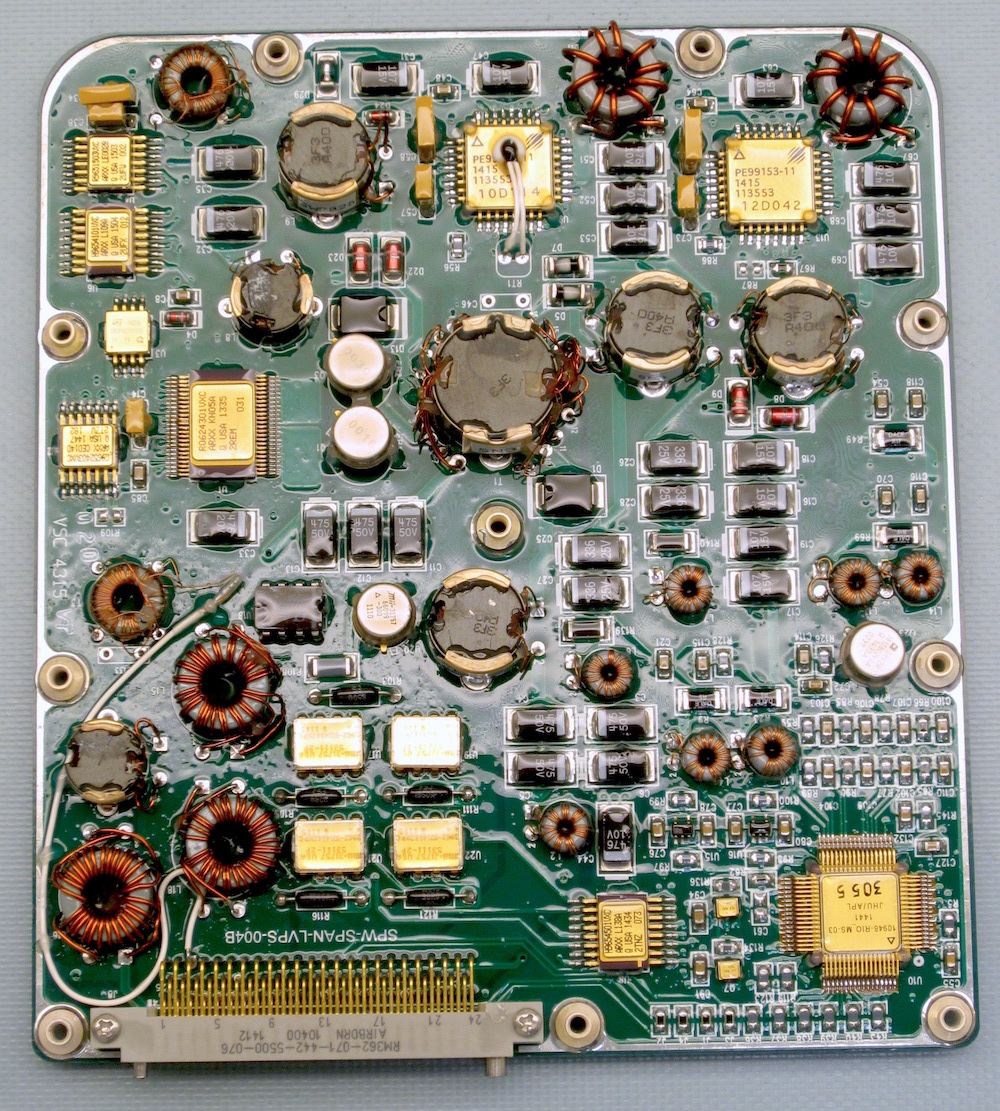} \includegraphics[height=5cm]{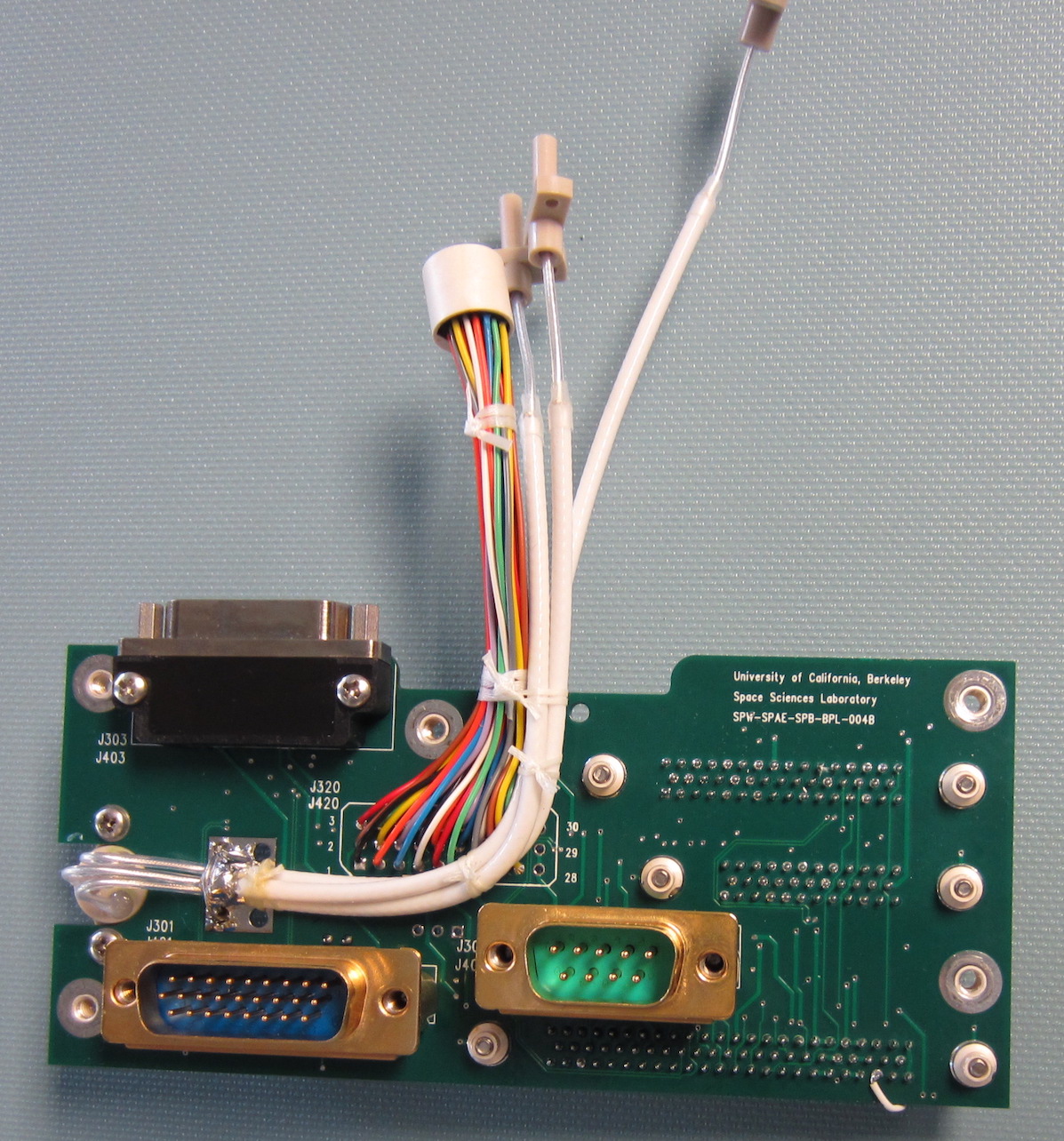} 
\caption{All of the electronics boards that are part of the electronics package for a single SPAN-E. Top Left: SPAN-Ae Anode board without MCPs or MCP mounting structure attached. The hole in the center is for the HVPS signal to pass through the boards below and reach the hemisphere. Top Center: SPAN-E Digital board, including FPGA and memory modules. The SPAN-E FPGA puts packets in CCSDS format for receipt by the SWEM. Top Right: SPAN-E HVPS, which creates hemisphere, deflector, MCP bias, and spoiler voltages, which are sent either though the center holes in the digital and anode boards, or through the backplane (Bottom Right). Bottom Left: SPAN-E LVPS, which generates 1.5V, 3.3V, 8V, and 5V for SPAN-E. Bottom Right: Backplane board for SPAN-E. The digital, HVPS, and LVPS are connected through the backplane. The connectors coming out of the top of the backplane supply signal to the attenuator and the high voltage signals to the spoiler and deflectors. The enable plug is not installed in the image.
\label{fig:boards_composite}}
\end{figure*}

Each SPAN-E instrument has its own electronics box that supplies low and high voltage to the instrument, accumulates counts, bins and packetizes data in Consultative Committee for Space Data Systems (CCSDS) format, and accepts commands from the SWEM. Each electronics package consists of four boards: an anode board that collects electron counts, a digital board that processes electron counts from the anode board into science data and performs digital processing, a high voltage power supply (HVPS) board that powers the analyzer optics and channel plates with high voltage (HV) signals, and a low voltage power supply board (LVPS) that generates voltages necessary for the other boards to operate. A fifth ``board,'' called the ``backplane,'' does not perform electrical tasks on its own, but rather acts as a connector between the digital, HVPS, and LVPS boards. It is also the location of the enable plug that permits or restricts power connection between the power supplies and the mechanical attenuator, as well as the hemispheres, deflectors, and the HVPS. A block diagram showing essential components of the SPAN-E electronics box and relevant boards is shown in Figure \ref{fig:block_diagram}, with pictures of the boards in Figure \ref{fig:boards_composite}.

\subsection{Anode Board}\label{sec:anode_board}

The SPAN-E anode board is mounted directly under the exit aperture of the analyzer optics. The SPAN-Ae and SPAN-B boards are electrically identical, with the exception of the layout and size of the anodes to accommodate differences in fields of view (see Section \ref{sec:mechanicalDesign}). The SPAN-E anode electronics serve the primary function of collecting electron charge clouds from the backside of the MCPs and converting them into digital signals. 

Metallized pads, called anodes, collect charge clouds from the back side of the MCPs and deliver the resulting current pulses through high voltage capacitors embedded in the board to the charge-sensitive preamplifier ASIC mounted on the back side of the anode board. Individual signal lines from each anode are separated from traces on either side of the board by multiple ground planes, preventing crosstalk between anode channels. The anodes span 240\textdegree of look directions on the anode board, and the MCPs span the same angle, albeit with a small amount of overlap.  

Another function of the anode board is to provide a structure for the MCP mounting fixtures. The SPAN-E MCPs are mounted in chevron pairs to increase the electron yield to millions of particles at the output. The plates were manufactured by Photonis and have a mean resistance of $46M\Omega$. 

The SPAN-E anode board has high voltage capacitors embedded into the layers of the board itself, which is a newmultilayered design compared to heritage sensors that allows for enough isolated capacitance to transmit signals through the anode board while maintaining a small footprint in the board layout. Since the back side of the MCP's are at $>$1kV, and we intend to accelerate the electrons from the back of the MCPs toward the anode board itself, electron ESAs typically require boards that are rated to several thousand volts. These high voltage capacitors effectively isolate HV signals, potentials, and circuitry from the low voltage electronics in the remainder of the instrument. In heritage instruments, the high voltage capacitors are placed on a separate dedicated board, adding mass and power consumption. With the inclusion of a 16-channel charge-sensitive preamplifier ASIC from the Laboratoire de Physique des Plasmas (LPP), it became possible to combine multiple electronics boards into one by installing the E-ASIC onto the low voltage side of the SPAN-E board, but only if the HV capacitors were also on the board \citep{Rhouni2012}. The embedded capacitors on the board vary in capacitance depending on anode size, but range between $60pF$ and $100pF$. 

\subsection{Digital Board}\label{sec:digital}
The SPAN-E digital board performs the main logic functions for the instrument, including executing commands, loading sweep tables, commanding the detection thresholds, and compiling the counts from the anode board into data, thereafter sending the packets to the SWEM for storage onboard or delivery to the spacecraft solid state recorder (SSR). It manages all command and telemetry signals to and from the SWEM.

SPAN-Ae and SPAN-B each have their own digital control board, allowing them to be commanded separately. Each SPAN-E can operate using different sweep tables, generating data products of different dimensions and cadences separately. At the time of publication, the SPAN-E sensors have been operated identically and simultaneously, with some minor exceptions during commissioning and during spacecraft orientations that place SPAN-B into direct sunlight (at which point it must be powered off due to thermal constraints). When the instruments are operated simultaneously and identically, it permits combining the two fields of view into one unified science product from synchronous measurements of electrons incident from all look directions.

An important function of the SPAN-E digital board is to store the discrete values in the instrument sweep tables onboard. Each SPAN-E has both rewritable MRAM and SRAM, which  contain the control values for the HV sweeping signals at any point during the sweep. More about the instrument measurement scheme is detailed in section \ref{sec:sweeping_meas}. 

The digital board houses the instrument's field programmable gate array (FPGA), which contains a processor, and sorts time-dependent signals from the anode board into the appropriate energy / deflector / anode bin in real time. All of the measurement acquisition timing originates from the FPGA, which itself runs on the processor clock signal sent from SWEM, which is often shared between FIELDS and SWEAP. The FPGA also controls the high voltage sweeps that are read from separate memory modules on the digital board, and calculates a rudimentary checksum on the tables to ensure that sweep tables are loaded from memory uncorrupted. It is also able to sum over multiple 218 millisecond acquisition periods to reduce the data rate and total volume generated by each instrument to ensure compatibility with the spacecraft SSR telemetry limits.


\subsection{High Voltage Power Supply Board}\label{sec:hvps}

The SPAN-E high voltage power supply board generates all of the high voltages that are used in the operation of the instrument. Maximum voltages for the instrument's different supplies are listed in Table \ref{tab:instrument_params}. The four sweeping supplies (hemisphere, deflectors, and spoiler) are each controlled by a digital to analog converter (DAC) chip, which converts the digital signal sent from the digital board (see Section \ref{sec:digital}) into an analog output that serves as the signal into the HV amplifier. For the hemisphere and spoiler supplies, that voltage reference is 4 volts for the entire range. For example, when the digital board commands the hemisphere supply (0-2000 V range) to go half-scale, a 2V signal is generated by the hemisphere's DAC, and the hemisphere supply biases the hemisphere to 1000V.

For the deflector supplies, however, the DAC is referenced relative to the hemisphere supply control voltage. Using this coupled DAC control voltage technique results in deflector biases scaled to the correct value for each hemisphere voltage step. 


\subsection{Low Voltage Power Supply Board}\label{sec:lvps}

The LVPS generates 1.5V, 3.3V, $\pm$ 5V, $\pm$ 8V secondary voltages from the 22V supplied by the SWEM. The SWEM-supplied 22V power source also passes through the backplane to the enable plug before reaching the digital and HVPS boards. The enable plug acts as a hardware failsafe to prevent high voltage from being applied to the analyzer surfaces in air or with the cover closed, preventing a damaging discharge event. The LVPS also reports the voltage and current draw on all of the supplies listed in this paragraph for inclusion in instrument housekeeping.


\subsection{Backplane Board}\label{sec:backplane}

The backplane board acts as a hardware bridge between the digital, HVPS, and LVPS boards. The anode and digital boards are connected to each other directly through a stacking connector. In addition, the backplane carries high voltage signals to the spoiler and deflectors via high voltage signals. The backplane also transmits command and telemetry signals from the cover opening mechanism and the mechanical attenuator between the digital board and the mechanisms. The enable plug on the SPAN-E instrument is mounted directly to the backplane. When absent, the enable plug also prevents the instrument cover and mechanical attenuator mechanisms from actuating, and directly provides a method to probe the LVPS voltages and currents.

\begin{deluxetable*}{c c c c c c}
\tablecaption{SPAN-E Instrument Sweep Modes from Encounters 1 \& 2
\label{tab:op_modes}}
\tablehead{
\colhead{\textbf{Mode Name}} & \colhead{\textbf{When Used}} & \colhead{\textbf{Energy Range}} & \colhead{\textbf{\# Energy Steps}}  & \colhead{\textbf{\# Deflector Steps}} & \colhead{\textbf{\# Anodes}}
}
\startdata
Nominal     & Encounters 1 \& 2 & 2eV-2keV    & 32 & 8\tablenotemark{b} & 16 \\
High Energy & Encounters 1 \& 2 & 10eV-10keV  & 32 & 8\tablenotemark{b} & 16 \\
\enddata
\tablenotetext{b}{Sweep tables used in Encounters 1 \& 2 included pre-launch deflector values, and as a result the outermost deflection angles in SPAN-E data are unreliable. Use caution when using the full 3D spectra when treating these outermost deflection $\theta$ angles. }
\end{deluxetable*}

\subsection{Measurement Operations} \label{sec:single_measurement}

The instrument is capable of producing 8 separate data packets, and each one is configurable in 5 ways: 1) archive or survey 2) product type and size 3) targeted or full (only applies to four of the eight products) 4) number of accumulation cycles acquired and 5) whether the product is summed over those accumulation cycles or whether a snapshot is taken every nth 218ms accumulation cycle. The only non-configurable aspect of all eight product packets is that half of the products are configured to enable measurement of ``targeted" spectra. Four products must always measure full spectra to select the peak signal bin for the targeted products to track, and the remaining four products have the option to be configured to measure full or targeted spectra.

When SPAN-E powers up, several configuration processes must occur before measurements can be made. First, on powerup, the instrument generates only housekeeping. Second, no sweep or product binning tables are loaded into instrument memory, so even if production of the eight science packets were enabled, they would contain no data. Hence, one of the first tasks of powering up SPAN-E is to load a set of sweep tables to control the analyzer high voltage surfaces, and also a set of product tables to sort the counts as a function of time into their appropriate energy / deflection / anode bins. Two product tables can be loaded at once, hence two different anode / deflector / energy bin dimension tables can be produced at the same time.

\begin{figure*}[ht!]
\centering
\includegraphics[height = 7.5cm]{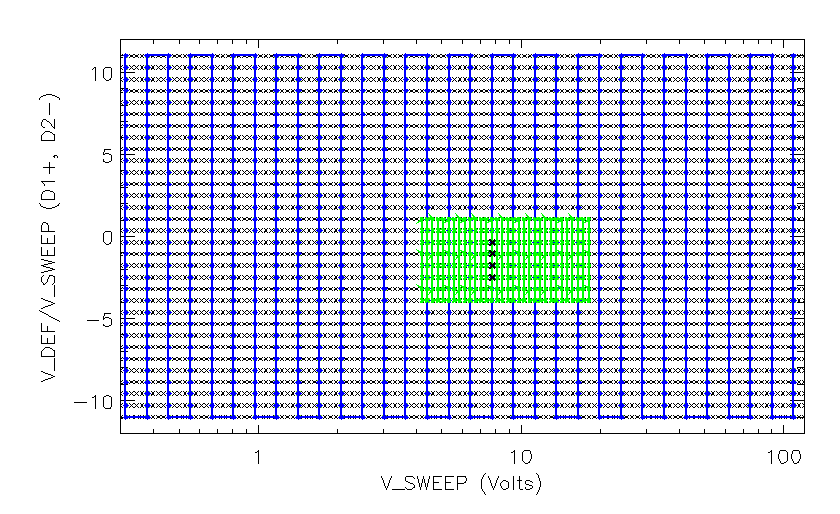}
\includegraphics[height = 7.5cm]{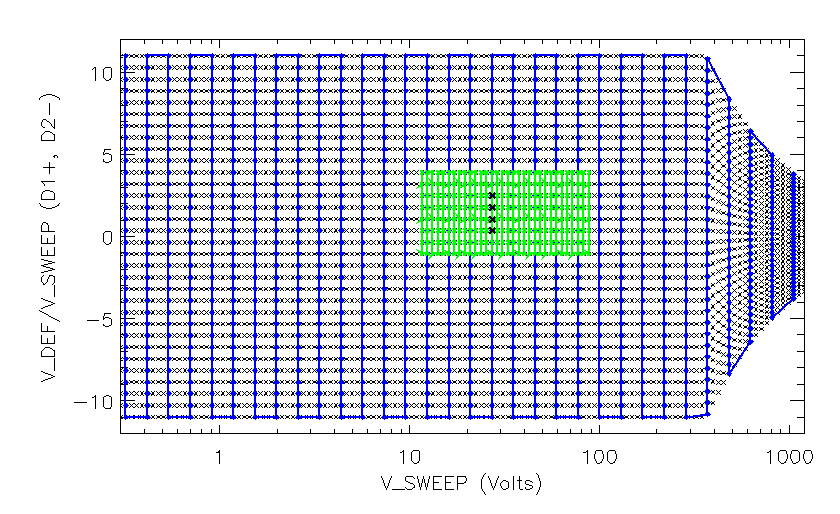}
\caption{A graphical representation of an entire sweep table for SPAN-E as loaded into instrument memory for two tables with different energy ranges. The black ``X" marks represent the entire range of possible deflector / hemisphere voltage combinations. The blue trace marks the combination and time order of the discrete steps in voltage values executed during a ``full" sweep; the green trace represents the values during a targeted sweep where the peak bin from the full sweep is indicated by the four bold black ``X"'s. Each red X is one of four ``microsteps'' in deflection, the counts from which are binned into a single deflection bin in the data products. The top figure is a table spanning the full FOV and $2eV - 2keV$, the bottom is the full FOV spanning $20eV - 20keV$. The Field of View is limited for measurements above 4keV due to the maximum allowable voltage on the deflector high voltage supplies.
\label{fig:fullAndTargSweep_2D}}
\end{figure*}

SPAN-E makes a complete measurement over 0.218 sec, scanning through 32 energy bins and 8 deflector bins during that time for Encounters 1 \& 2. The instrument accumulates counts for the duration of one 32 energy $\times$ 8 deflector phase space bin, which occurs for 0.850 milliseconds. After making one complete measurement cycle (0.218 sec) over the full energy range defined in the currently loaded sweep table, the next measurement cycle will sweep over a reduced energy and deflection range with higher resolution targeted on the energy / deflection bin that contained the largest number of counts. This sweep, called the ``targeted" sweep, alternates every other measurement cycle with the ``full" sweep. Using this alternating measurement technique, SPAN-E can measure both a wide range of electron energies and flow directions, as well as measure a subset of the energy range with a higher resolution to resolve peaks in the distributions. The energy / deflector bin that the targeted sweep focuses on can be manually commanded, or permitted to track the bin with the largest signal. A graphical representation of the sweep tables relative to each other is found in Figure \ref{fig:fullAndTargSweep_2D}.

After the counts are collected on the Anode board, the signals are sorted on the digital board into the appropriate time-ordered product table. Each SPAN-E can have two products generated, which are themselves accumulated for an individually configurable number of 218 millisecond acquisition periods. Lastly, the tracked peak energy / deflector bin is fixed during multiple sequential summed acquisition periods in order to prevent the targeted spectra products from having ``mixed" energy and deflector sense; if the tracked bin were permitted to vary between targeted sweeping periods, then different energy bins would be summed over a long time and the measurement would be indeterminate.


\section{Instrument Operation} \label{sec:instrument_operation}


\subsection{Making a Measurement: Timeline}\label{sec:sweeping_meas}

SPAN-E by default begins its sweep at the highest hemisphere voltage in the sequence, thereafter stepping down in discrete increments for the remaining 31 voltage steps. For each hemisphere voltage step, the deflector supply will sweep through 8 steps times 4 microsteps (during full sweeps). In this case, a microstep is a voltage step on the deflector that does not have its own separate accumulation period. The default is to bias the upper deflector first, then after the deflectors complete one scan in $\theta$, the hemisphere steps down in voltage one step and the deflectors sweep through the opposite direction in $\theta$, alternating thereafter until the lowest step of the sweep table is reached. 

\begin{figure*}[ht!] 

\centering
\includegraphics[height = 6.5cm]{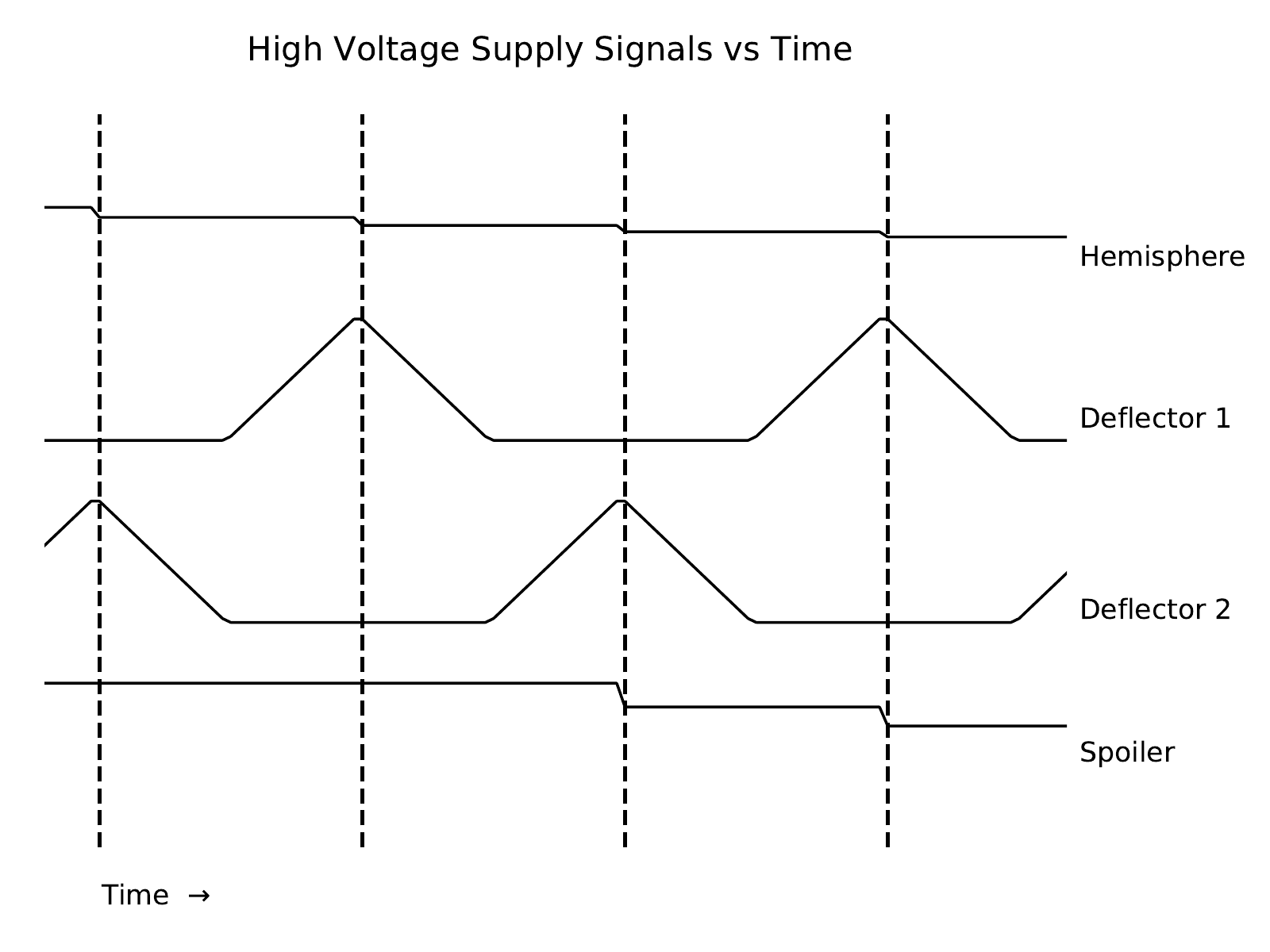}\includegraphics[height = 6.5cm]{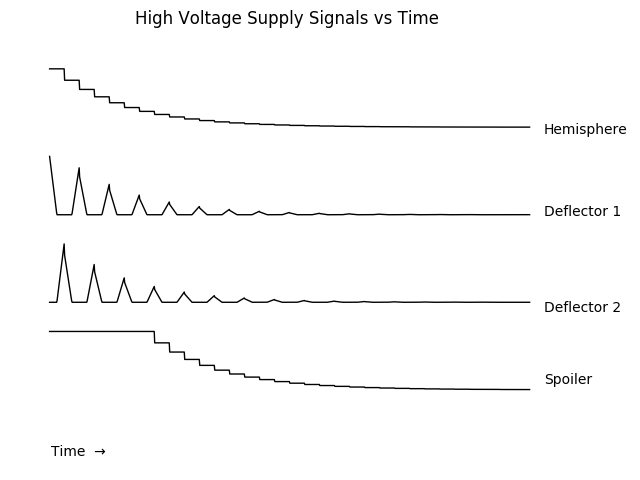}
\caption{Left, A cartoon image representing the four high voltage supplies that select electrons in SPAN-E as a function of time (arbitrary) for a full energy and deflector range sweep. From top to bottom, the hemisphere (select for energy), deflectors 1 \& 2 (select for theta direction) and spoiler (attenuate by energy). The signals in the above plot have been normalized and skewed by a constant value relative to one another, hence no axis labels are provided. Vertical dashed lines represent changes in the hemisphere value, during which time the deflectors will scan through $\theta$ in discrete commanded steps as stored in the instrument tables. The spoiler changes value at the same time as the hemisphere, thus electrons of a specific energy are attenuated with a constant factor of attenuation in all look directions below 500eV. Right, A typical full sweep pattern showing voltages from the four swept HV supplies in SPAN-E. The signals are offset and normalized to their maximum values to show their relative timing. Note that the hemisphere supply sweeps from a high value to a low one: in other words, SPAN-E scans high energies in an individual sweep before low energies. The spoiler supply is held constant at high energies, and steps down in voltage synchronous with the hemisphere at lower hemisphere voltages. The deflector supplies alternate which is biased and which is held at ground.
\label{fig:sweep_pattern}}
\end{figure*}

If the electrostatic spoiler is engaged, the spoiler steps down synchronous with the hemisphere. The spoiler power supply has a maximum of 80 volts applied. When the hemisphere is higher than the spoiler supply maximum, the spoiler voltage is held constant at the max value for a reduced factor of attenuation for electrons above $\sim$ 1200eV. See Figure \ref{fig:sweep_pattern} for a graphical representation of the order of sweeps in SPAN-E. 

The SPAN-E sensors make a complete acquisition over the full range of energies and deflections in 0.218 milliseconds. The nominal energy tables that are being used to sweep the hemispheres, deflectors, and spoiler span a range sufficiently large that there are gaps in coverage compared to the instrument's intrinsic $\Delta E/E$. After a full sweep is complete, the energy / deflector / anode bin with the largest signal is saved, and a targeted sweep around that energy / deflector step is initiated (see Section \ref{sec:instrument_operation}). The SPAN-E instrument initiates a targeted energy and deflection range sweep of finer resolution at the instrument's intrinsic energy and angular resolution in order to cover gaps in energy or deflection. This "targeted" sweep is initiated every other 0.218 seconds unless the instrument is specifically configured otherwise. This targeted acquisition is centered around the energy/deflector bin with the peak signal from the previous full sweep acquisition period. The order of the swept supplies, eg, the deflector scanning one direction through $\theta$ and then alternating, the hemisphere stepping down in voltage, etc, is the same in the targeted sweep as the full sweep, the only difference being that the deflector steps through 8 $\theta$ values instead of 8 * 4 steps and microsteps.

\begin{deluxetable*}{l c c c c c c c c}
\tablecaption{SPAN-E Data Acquisition Modes from Encounters 1 \& 2: Level 1 \& 2 Data 
\label{tab:data_modes}}
\tablehead{
\colhead{\textbf{Data}} & \colhead{\textbf{When}} & \colhead{\textbf{Product}} & \colhead{\textbf{Product}} & \colhead{\textbf{Cadence}} & \colhead{\textbf{Integration}}& \colhead{\textbf{Anode}} & \colhead{\textbf{Deflection}} & \colhead{\textbf{Energy}} \\[-8pt]
\colhead{\textbf{Type}\tablenotemark{a}} & \colhead{Used} & \colhead{\textbf{Type}} & \colhead{\textbf{Name}} & \colhead{\textbf{(sec)}} & \colhead{\textbf{Time (sec)}} & \colhead{\textbf{Bins}} & \colhead{\textbf{Bins}} & \colhead{\textbf{Bins}}
}
\startdata
SF0\tablenotemark{b} &              & 3D Spectra       & 16A$\times$8D$\times$32E & 27.96 & 13.98 & 16 & 8 & 32 \\
SF1 & Encounter 1  & Energy Spectra   & 32E                      & 1.748 & 0.873 &  - & - & 32 \\
AF0 & (Default)    & 3D Spectra       & 16A$\times$8D$\times$32E & 0.873 & 0.436 & 16 & 8 & 32 \\
AF1 &              & Energy Spectra   & 32E                      & 0.436 & 0.218 &  - & - & 32 \\
\\
SF0 &              & 3D Spectra       & 16A$\times$8D$\times$32E & 27.96 & 13.98 & 16 & 8 & 32 \\
SF1 & Encounter 1  & Anode Spectra    & 16A                      & 1.748 & 0.873 & 16 & - & -  \\
AF0 & (Diagnostic) & 3D Spectra       & 16A$\times$8D$\times$32E & 0.873 & 0.436 & 16 & 8 & 32 \\
AF1 &              & Anode Spectra    & 16A                      & 0.436 & 0.218 & 16 & - & -  \\ 
\\
SF0 &              & 3D Spectra       & 16A$\times$8D$\times$32E & 13.98 & 6.965 & 16 & 8 & 32 \\
SF1 & Encounter 2  & Energy Spectra   & 32E                      & 1.748 & 0.873 & -  & - & 32 \\
AF0 &  (Default)   & 3D Spectra       & 16A$\times$8D$\times$32E & 0.873 & 0.436 & 16 & 8 & 32 \\
AF1 &              & Energy Spectra   & 32E                      & 0.436 & 0.218 & -  & - & 32 \\ 
\\
SF0 &              & 3D Spectra       & 16A$\times$8D$\times$32E & 895   & 13.98 & 16 & 8 & 32 \\
SF1 & Cruise Mode  & Energy Spectra   & 32E                      & 112   & 13.98 & -  & - & 32 \\
AF0 &  (Default)   & 3D Spectra       & 16A$\times$8D$\times$32E & 27.86 & 13.98 & 16 & 8 & 32 \\
AF1 &              & Energy Spectra   & 32E                      & 27.86 & 13.98 & -  & - & 32 \\ 
\enddata
\tablenotetext{a}{Targeted sweep products are not included in this table, but have identical formats to their full counterparts; ``SF0'' is a ``Full" energy range product, and ``ST0'' is its ``Targeted'' range counterpart}
\tablenotetext{b}{``S'' stands for ``Survey'', ``A'' stands for ``Archive'', ``F'' stands for ``Full'', ``T'' stands for ``Targeted''.}
\end{deluxetable*}


\subsection{On-orbit Operation}\label{sec:ops_orbit}

Operation of SPAN-E can be divided into two main modes based on spacecraft operations: the primary ``encounter" phase of the orbits, which is approximately ten days centered around PSP perihelion (variable by orbit profile), and the rest of the orbit, hereafter called ``cruise" phase. The instrument measurement rate is higher and uninterrupted during nominal encounter phases as compared to cruise phases, during which the data rate is considerably lower and the measurement periods are interrupted by spacecraft communications, power limitations, and other spacecraft critical operations. During periods of interest (e.g., Venus encounters), the sensors can be configured to collect more data than typical outside of encounter. 

The SPAN-E operational and data acquisition modes during the first two solar encounters are summarized in Tables \ref{tab:op_modes} and \ref{tab:data_modes}.


\section{Data Description} \label{sec:data_description}

Data products generated from SPAN-Electron are classified according to the level of calibration required to produce the files, and the data type that is contained in the file.  Data files are produced in the Common Data Format (CDF) files, available at  https://sweap.cfa.harvard.edu/pub/data/, and archived in NASA's Space Physics Data Facility (SPDF). 


\subsection{Level 0 Data}\label{sec:L0_data}

Level 0 (L0) files are unprocessed files downlinked directly from PSP through the Deep Space Network (DSN) in their original packetized format created by the spacecraft. Files contain a fixed volume of data, and are named based on their date of acquisition. On their own, Level 0 files are not useful for scientific analysis, but are archived for troubleshooting purposes.


\subsection{Level 1 Data}\label{sec:L1_data}

Level 1 (L1) files are converted from the binary L0 format into a format readable by a standard data processing environment, such as IDL or Python packages. The SWEAP team uses either IDL or Python routines in the data production pipeline (depending on the instrument) to convert L0 files into L1 CDF files. To produce L1 files, minimal processing is performed since the intention of the L1 data is to serve as an archive of the instrument performance in its most raw state. All quantities in the CDF files are in engineering units, e.g. particle counts per accumulation period per energy bin number, deflection bin number, and anode number; physical units and coordinates $E (eV)$, $\phi$ (\textdegree), and $\theta$ (\textdegree) are absent. Because of the units, L1 files are not useful for scientific analysis. The intention behind archiving L1 files is to keep a record independent from scientific conversions for pipeline debugging purposes and instrument calibration consistency checks over the course of the mission. Housekeeping values are converted into temperatures, currents, and voltages. L1 files are available by request.


\subsection{Level 2 Data}\label{sec:L2_data}

Level 2 (L2) data files are generated from L1 files. Instrument units are converted into physical units. For example, counts per accumulation period are converted into differential energy flux as a function of energy in electron-Volts (eV), and deflection and anode bin numbers into degrees in $\phi$ or $\theta$. The L1 coordinates are still in the instrument frame of reference - in this regard the coordinate systems between SPAN-Ae and SPAN-B will differ. Level 2 data are released to the public for scientific analysis. 

Level 2 filenames directly reflect the type of data contained in the CDF. For example, to look at the full 3D spectra from SPAN-Ae on November 4th, 2018 (a day before the first perihelion), one would look at Table \ref{tab:data_modes} and see that for the first encounter the ``SF0" product is the appropriate product to look at, which is a component of the L2 filename. Likewise, each CDF L2 data file name contains name of the product (``Product Name'' in Table \ref{tab:data_modes}) for a given data type (``Data Type'' in Table \ref{tab:data_modes}). The L2 data is separated into individual day long files, such that each file contains all the individual sensor data for the data specified in its filename.


\subsection{Level 3 Data}\label{sec:L3_data}

Level 3 (L3) products are, by definition, any data product which requires information from another source to produce, such as magnetic field data (from FIELDS), or data from another sensor on SWEAP. Likewise, any functions performed on L2 files including generation of moments or other processing which expands or reduces the number of dimensions of the L2 data are produced as a L3 product. Notably, for SPAN-E, the L3 data specification means that a combined data product from both sensors is a L3 product, so a full-sky 3D distribution combining SPAN-Ae and SPAN-B measurements would be a L3 product. Other L3 products include electron pitch angle distributions (PADs), which are processed using the FIELDS magnetic field L2 data. Electron moments and fits, which produce values of density, temperature, and velocity, would also be L3 products since they are created through combinations and modifications of the L2 data.


\section{Observed Performance on Orbit \#1 and \#2}\label{sec:observed_performance}

\subsection{Overview of Encounters 1 \& 2}\label{sec:overviewE1E2}

\begin{figure*}[ht!] 
\centering
\includegraphics[height = 12cm]{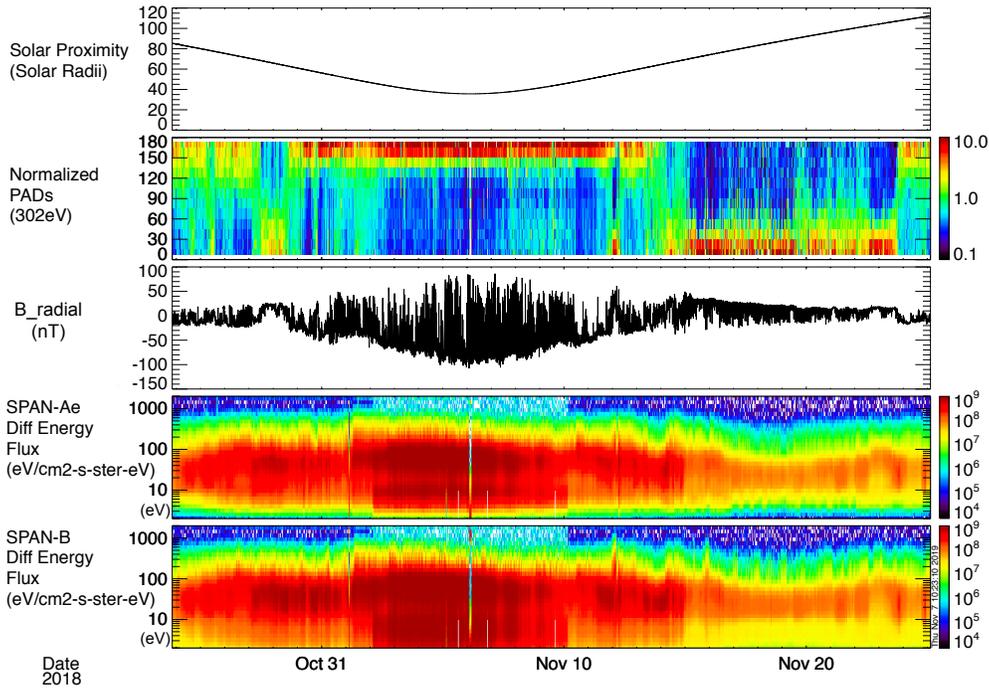}
\caption{The first PSP encounter as a function of time. Top to bottom: PSP distance from the solar center sun in solar radii; combined SPAN-E PADs for the passband at 302eV and normalized by the average flux; the R component of the ambient magnetic field in a heliocentric reference frame in nT; summed energy spectra (SF1) from SPAN-Ae in differential energy flux units; summed energy spectra from SPAN-B in differential energy flux units. During most of the first encounter, the magnetic field is pointed sunward, and hence the PADs for the 302 eV channel that corresponds to the sense of the electron strahl is found at 180\textdegree.
\label{fig:E1}}
\end{figure*}

The SPAN-E instruments performed nominally in Encounters 1 and 2, taking nearly 80,000 3D measurements of the electron distribution functions per sensor. An overview of the data taken by SPAN-Ae and SPAN-B during encounter 1 are shown in Figure \ref{fig:E1}, and the operational modes used by the instrument during those encounters is found in Table \ref{tab:data_modes}.


\subsection{Known Instrument Caveats}\label{sec:datacaveats}
\begin{figure*}[ht!] 
\centering
\includegraphics[height = 7.0cm]{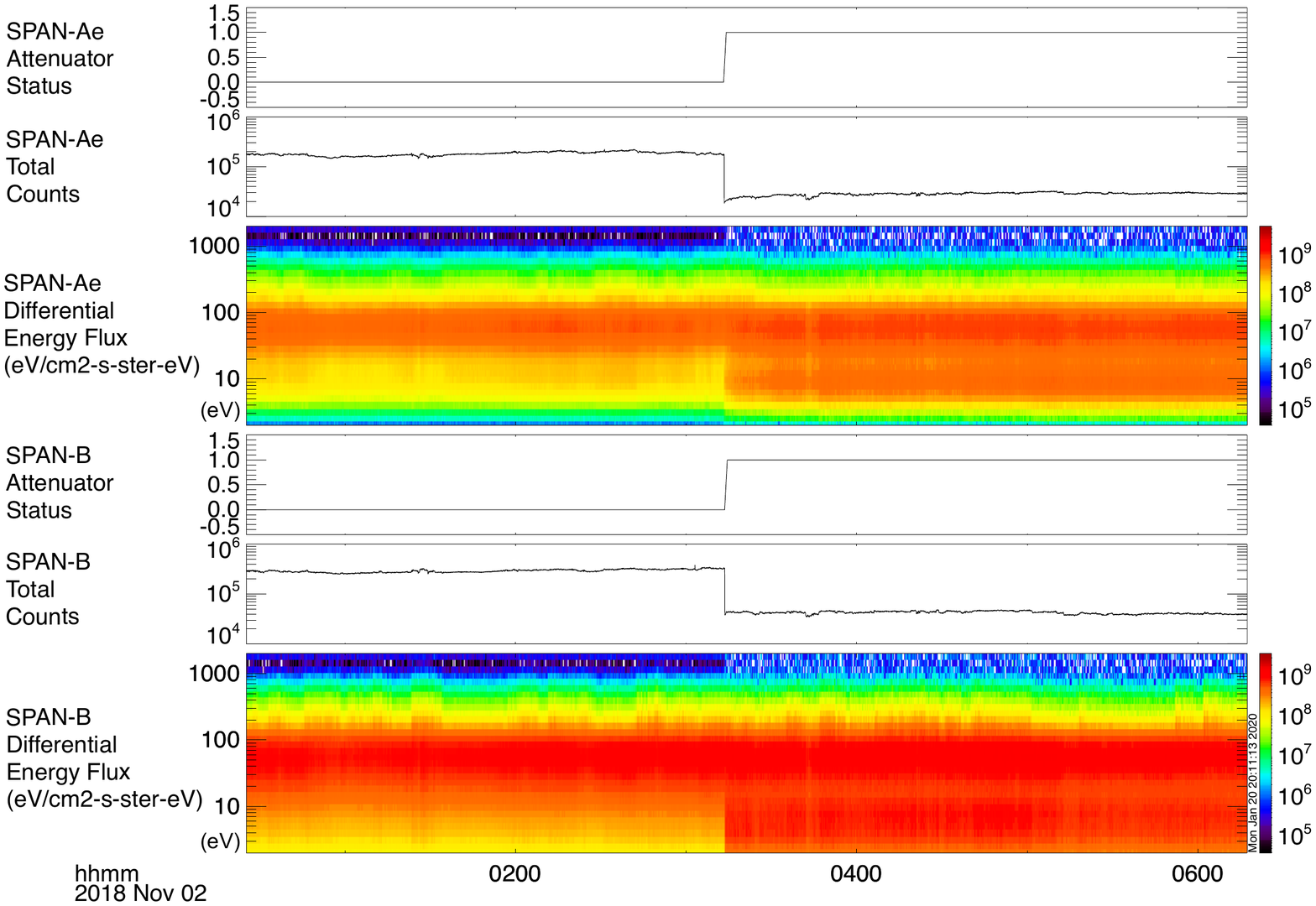}\includegraphics[height = 7.1cm]{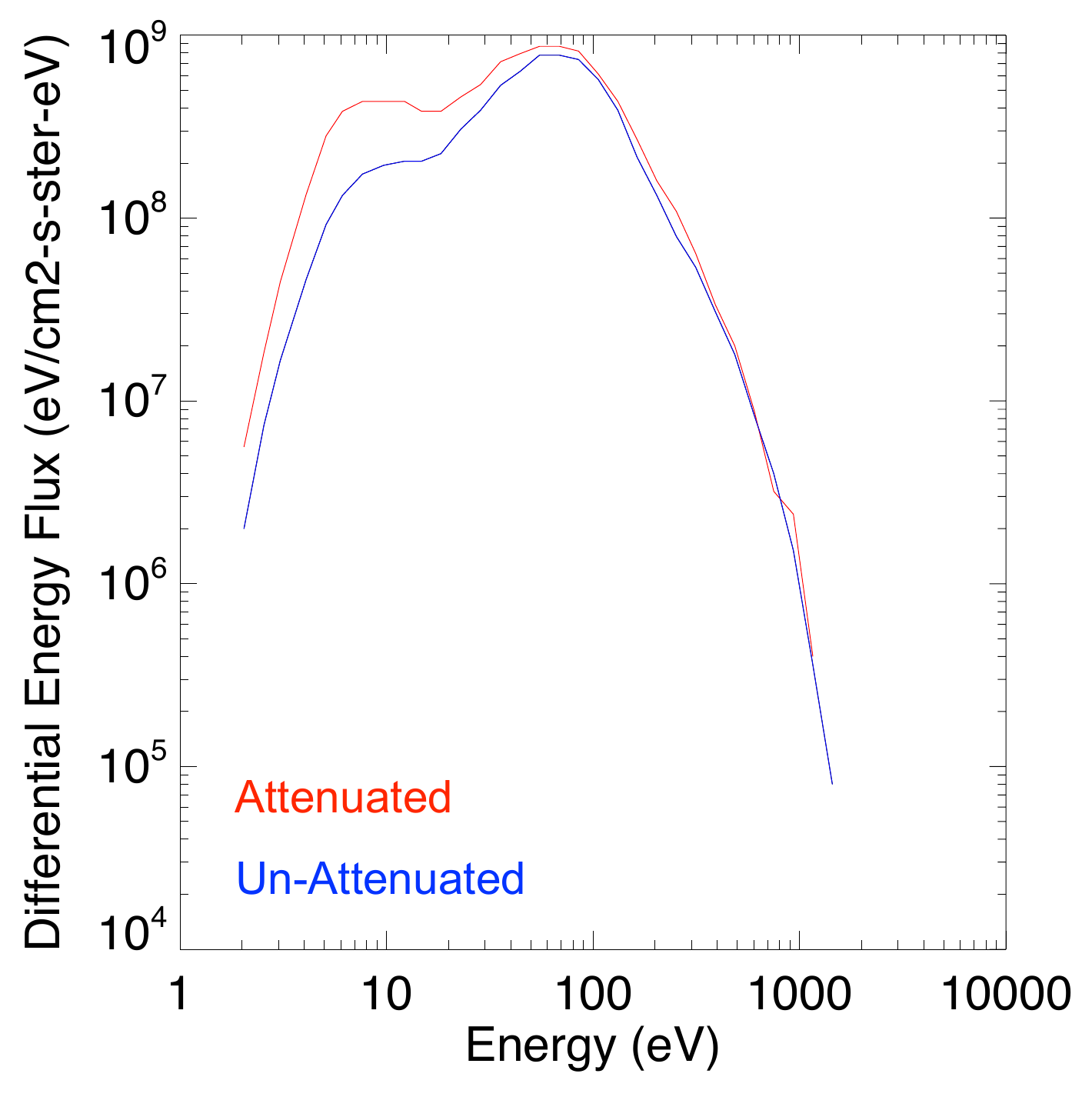}
\caption{\textbf{Left:} Time-series plot of the SPAN-E mechanical attenuators being engaged in the first encounter. Top to Bottom: Attenuator Status, SPAN-Ae, Total Counts per accumulation period, SPAN-Ae (generated onboard instrument), Summed Energy Spectra for SPAN-Ae, Attenuator Status, SPAN-B, Total Counts per accumulation period, SPAN-B (generated onboard instrument), Summed Energy Spectra for SPAN-B. The attenuator moves to the ``In'' position at approximately 0310, with ground software constructing an artificial increase in ``background'' electron flux thereafter. \textbf{Right:} single energy spectrum produced by SPAN-Ae before (blue) and after (red) the mechanical attenuator is engaged (same time period). An artificial increase in electron flux below approximately 25 eV is apparent.
\label{fig:atten_in}}
\end{figure*}

The SPAN-E instruments, although performing as expected and designed in Encounters 1 and 2, are imperfect sensors, and thus there are a few caveats to be considered for anyone looking at SPAN-E data at the time of this publication. These caveats are expected to be addressed in future revisions of the CDF data files, but for now they are stated here.

\subsection{Caveat: Mechanical Attenuator and the Halo Electrons}

The SPAN-E mechanical attenuator (see Section \ref{sec:attenuators} was manually commanded to be inserted into the field of view of both sensors four days before Perihelia 1 and 2, and was removed via command four days after those perihelia. In the L2 data processing, a constant factor of 10 is used to calculate the differential energy flux during the periods when the attenuator is inserted. For lower energy data bins, where the electron flux is comparatively high, this works as expected. However, for higher energy bins where the halo electrons are less abundant, the instrument background counts from the microchannel plates begins to be comparable to the electron halo measurement. Thus, when the constant factor of 10 is applied to calculate differential energy flux, the background noise in energy bins where halo electrons are measured is also multiplied by a factor of 10. Thus, during the eight-day period surrounding perihelia in Encounter 1 and Encounter 2, any attempt to fit the halo must take into account an increase in effective instrument noise as well.

Additionally, due to the mechanical attenuator acting as a metallic partial obstruction to the instrument field of view, the secondary electron production off of surfaces inside the instrument increases when the mechanical attenuator is engaged. This can be seen in figure \ref{fig:atten_in} and is detailed in the next section.

\subsection{Caveat: Ambient Secondary Electron Production}

Secondary electron production, or, electrons that are generated as a result of a primary electron impact into material, is fairly efficient for primary impacting electrons with energy above 150eV or so. Due to the hot temperature of the electrons in the inner heliosphere, many secondary electrons are created due to impact from primaries above 50 eV or so. In the PSP electron data, these secondaries contribute significantly to the total measured electron flux for measurements below $\sim 25$ eV. At the time of this publication, it is not recommended to use electron data below $\sim 25$ eV in the Level 2 files. A focused calibration effort to quantify the secondary electron populations' contribution to energy flux is underway, and results will be included in a future SPAN-E instrument calibration paper.

Electron measurements over the entire energy range are provided in the L2 CDF files. However, in the L3 PADs, energies below $\sim 25$eV are not included in the L3 PAD CDFs because they are still contaminated by ambient and mechanical attenuator generated secondary electrons.


\subsection{Caveat: Field of View Obstructions}

\begin{figure*}[ht!] 
\centering
\includegraphics[height = 7.5cm]{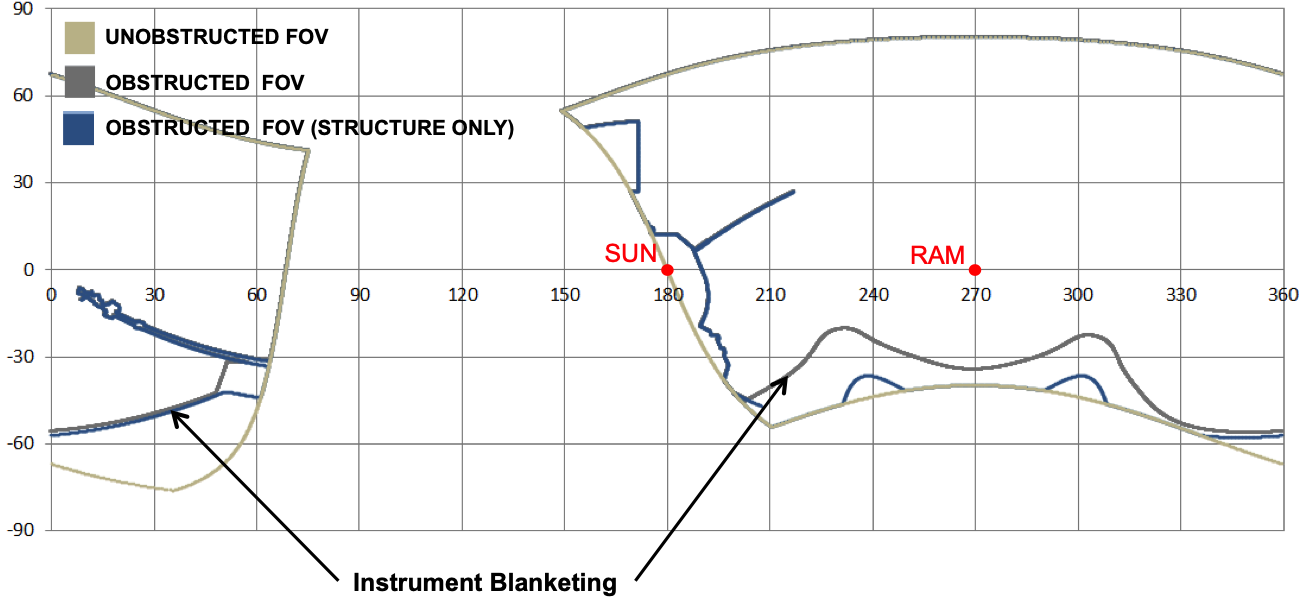}
\caption{Obstructions to SPAN-Ae's field of view. The instrument's full, unobstructed field of view is contained inside the beige rectangular area, with un-blanketed spacecraft intrusions to the FOV show in blue. The mag boom is seen at the left (10-60\textdegree, -15\textdegree) and the heat shield and FIELDS Antenna are seen radiating from the sunward direction (center). At (165\textdegree, 40\textdegree) the solar panel is indicated. An estimate of the actual blockage, thermal blankets included, is shown in grey.
\label{fig:spanae_blockage}}
\end{figure*}

\begin{figure*}[ht!] 
\centering
\includegraphics[height = 7.5cm]{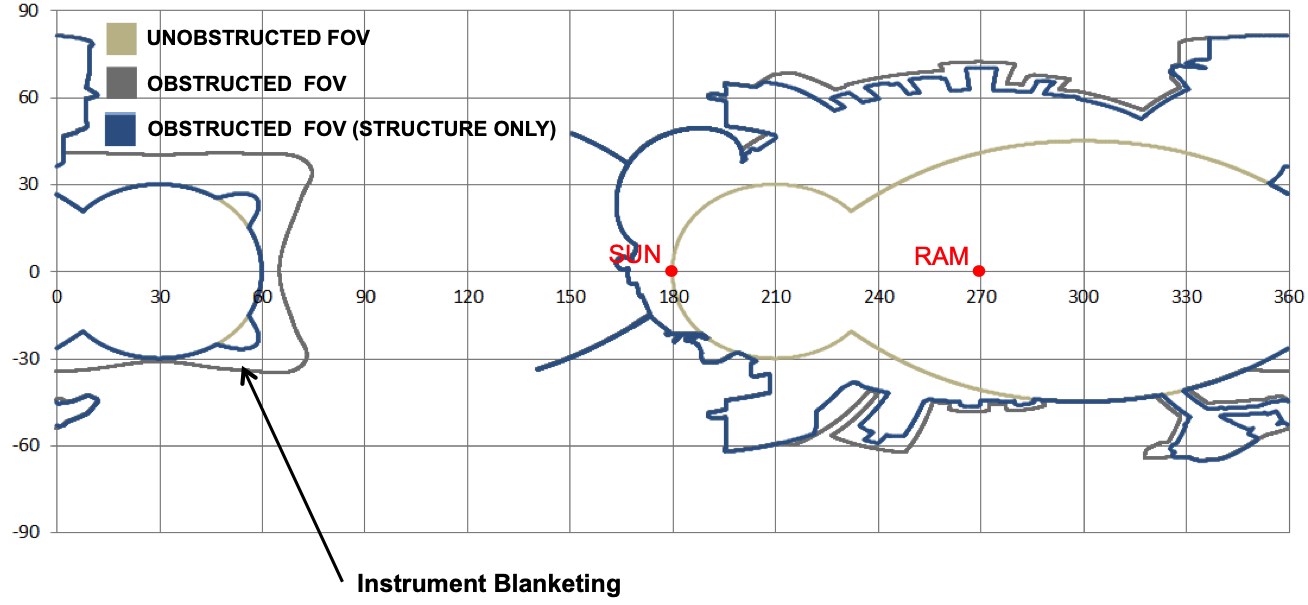}
\caption{Obstructions to SPAN-B's field of view. The inner boundary of the instrument's full, unobstructed field of view is shown by the beige boundary, within which the instrument does not measure electrons. The spacecraft intrusions to the SPAN-B FOV are substantially greater than SPAN-Ae, and are shown in blue, including sun sensors, solar panels, fuel filling ports, FIELDS antennas, hi-gain antenna, and more. An estimate of the actual blockage, thermal blankets included, is shown in grey.
\label{fig:spanb_blockage}}
\end{figure*}

The nature of the 3-axis stabilized PSP spacecraft necessitates use of multiple ESAs in order to get a near full sky view of the solar wind electrons. However, the small size of the spacecraft combined with numerous necessary instrumentation additions for operating the spacecraft meant that it was impossible to preserve the full fields of view of both SPANS while still keeping them safely in the umbra of the heat shield. Thus, the spacecraft team attempted to keep as much of the fields of view unobstructed while also making space for critical additions such as the sun sensors and, of course, the heat shield. The end result is estimated in figures \ref{fig:spanae_blockage} and \ref{fig:spanb_blockage}, which combine the known solid blockages as generated from the spacecraft model with estimates of thermal blanketing thickness from all over the spacecraft on a full-sky map. These can be compared to the full sky map in figure \ref{fig:FOV}.

\subsection{Caveat: High Voltage Supply Hysteresis}

\begin{figure*}[ht!] 
\centering
\includegraphics[height = 12cm]{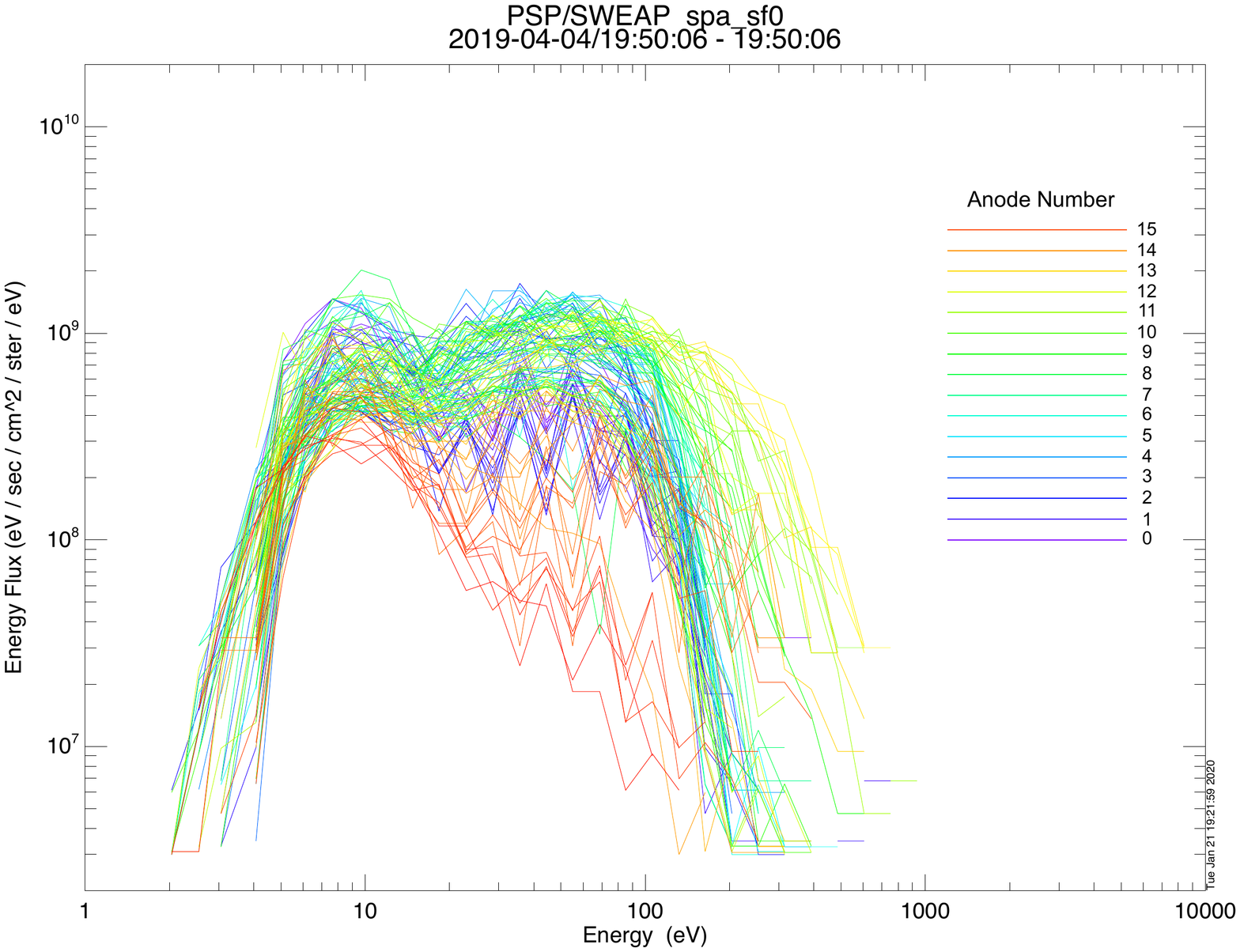}
\caption{A plot of each angular (phi by theta) bin response as a function of energy. The sawtooth pattern seen in the red and purple corresponds to extreme theta deflection angles and outer phi pixels, and indicate a difference in measurement with different deflection sweeping direction.
\label{fig:hysteresis}}
\end{figure*}

The SPAN-E sensors step through their four high voltage supplies as a function of time, as indicated in section \ref{sec:hvps} and \ref{sec:single_measurement} and \ref{sec:measurement_electronics}. For adjacent energy steps, the deflector sweeps from negative to positive theta angles in one step, switching from positive to negative angle steps in the other direction. This alternating sweep pattern is intended to save supply power and prevent retrace from contaminating the measurements. However, as can be seen in figure \ref{fig:hysteresis}, the response is not equal at all energies, indicating differing behavior in the instrument when deflectors are sweeping negative to positive theta vs. positive to negative theta. Although the definitive source of this anomaly is still under investigation, a prime suspect is a hysteresis between the expected voltage on the deflector supplies and the actual value. This could be due to a potential time delay between the commanded deflector supply voltage and the actual time of that supply reaching the commanded voltage, or different DC offsets between the upper and lower deflector high voltage supplies. 

Regardless of the root cause, the effect remains uncorrected in the public SPAN-E L2 data released at the time of this publication. The effect is seen primarily at extreme deflection angles, namely those outermost theta bins. Therefore, at this time the user is advised to not use the outermost deflection bins without consulting the SWEAP team for advice on how to proceed. 


\section{Conclusion}

The SPAN-E sensors will make critical measurements of solar wind electrons at the acceleration and heating regions in the inner heliosphere and upper corona. The sensors are a high-heritage, matched pair of toroidal top-hat electrostatic analyzers with a large range of deflection. Together these two sensors will characterize the low energy electron populations from all look directions on the PSP spacecraft. Their fields of view have been optimized for making high resolution measurements of the electron strahl width, and the energy ranges are fully adjustable to adapt to the predicted changing plasma environment over PSP's seven year long mission. The SPAN-E measurement of strahl electrons will provide topological context for the large scale structures, such as Coronal Mass Ejections and flux ropes, that are expected observations over the duration of the Parker Solar Probe mission. By fully characterizing the strahl, core, and halo electron sub-populations as functions of energy and direction, the SPAN-E sensors are providing key measurements for science questions that address plasma instabilities and microphysics as they pertain to coronal heating and solar wind acceleration, which are two key points in the Parker Solar Probe driving science goals. Working in tandem with the whole suite of the plasma instrumentation on PSP, the SPAN-E sensors play a key and complementary role in unraveling the mysteries of the solar corona by fully characterizing the plasma environment closer to the sun than ever before.

\section{\textbf{Acknowledgements}}
\acknowledgments

\textit{This work was funded by NASA contract NNN06AA01C.  The authors wish to acknowledge the significant work of all of the technical and engineering staff that worked on the spacecraft and SPAN-Electron instrument, especially Chris Scholz, Matt Reinhart, Andrew Peddie, and Dave Mitchell for their invaluable suggestions on the manuscript. Thanks also to CNES for providing the charge-sensitive preamplifier ASICs which were critical components of this instrument design. Lastly, many thanks to the reviewer for their detailed comments and suggestions that the authors feel have elevated the quality of the manuscript.}

\bibliography{references}

\end{document}